\documentclass{mn2e}
\usepackage{graphicx}
\newcommand{\bq}{\begin{equation}}
\newcommand{\eq}{\end{equation}}
\newcommand{\bqa}{\begin{eqnarray}}
\newcommand{\eqa}{\end{eqnarray}}

\begin{document}

\title{Non-adiabatic Oscillations of Red Giants}
\author[D.R. Xiong \& L. Deng]{D.R. Xiong$^1$ \& L. Deng$^2$\\
$^1$Purple Mountain Observatory, Chinese Academy of Sciences, Nanjing 210008; Xiongdr@pmo.ac.cn\\
$^2$National Astronomical Observatories, Chinese Academy of
Sciences, Beijing 100012; licai@bao.ac.cn}

\maketitle

\begin{abstract}
Using our non-local time-dependent theory of convection, the
linear non-adiabatic oscillations of 10 evolutionary model series with
masses of 1--3M$_\odot$ are calculated. The results
show that there is a red giant instability strip in the lower
temperature side of the Hertzsprung-Russel (HR) diagram which goes
along the sequences of the red giant branch (RGB) and the asymptotic
giant branch (AGB). For red giants of lower luminosities,
pulsation instability are found at high order overtones, the lower order
modes from the fundamental to the second overtone are stable.
Towards higher luminosity and lower effective
temperature, instability moves to lower order modes, and the
amplitude growth rate of oscillations also grows. At the high luminosity end of
the strip, the fundamental and the first overtone become unstable,
while all the modes above the 4th order become stable. The excitation
mechanism have been sdudied in detail. It is found that turbulent
pressure plays a key role for excitating of red variables. The frozen
convecttin approximation is unavailaable for the low temperature stars
with extended convective envelopes. In any case, this approximation can
explain neither the red edge of the Cepheid instability strip, nor the
blue edge of the pulsating red giant instability strip. An analytic
expression of a pulsation constant as a function of stellar mass,
luminosity and effective temperature is presented from this work.
\end{abstract}

\begin{keywords}
convection---stars:variables---stars:late-type---stars:evolution
\end{keywords}

\section{Introduction}
The Mira type variables are among the most luminous stars in a star
cluster, and they possess very large light amplitudes
therefore they are the first variable stars ever discovered. The red
variables, however, are not yet understood well. According
to Wood (2000) "red giant stars are probably the least understood
of all variable stars". It was suspicioned long time ago that all red stars are
variables. Eggen (1969) pointed out, for instance, that all old disc
population red giants with (R-I)$_0\geq 1.0$ are variables. A few
year later, he shifted the blue edge of that instability area to
(R-I)$_0\approx 0.9$ (Eggen 1973). The visual amplitudes of red variables
have vary large ranges, from 0.01 to 10 magnitudes, and the typical time
scales (periods) goes from several to a few
hundred days. Classification of these red variables based on period,
luminosity and population data were made by Eggen (1972a,b, 1973a,b,
1975, 1977). By photometric and spectroscopic observations of 187 G
\& K type giants, Henry et al. (2000) showed that all
K5-M0 type giants are variables, and they further claimed that
the variations of red giants with spectral type later than K2
are resulted from pulsation. Towards lower effective
temperature and higher luminosity, the light amplitude
tends to increase. In fact, these stars are located either in the
upper part of RGB or on AGB. It is identified by Percy et al. (1998,
2001, 2003) that these stars are actually pulsating from the
fundamental up to the 3rd overtone. It is also found that there are
variables with rather short periods from several hours to $\sim$2 days
and small amplitudes from several to 20 percents of a magnitude
on the lower RGB of globular clusters (Yao \& Qin
1993, Yao, Zhang \& Qin 1993). By analyzing the 38.5 day photometric
data acquired by HST for 47 Tuc, Edmods \& Glliland (1996)
discovered a mount of small amplitude K type variables on RGB, whose
periods go from 2 to 4 days and the amplitudes only measure
5--15mmag. On the V--(B-V) color magnitude diagram (CMD) of 47 Tuc,
one group of these variables gathers near the red edge of the horizontal branch
(HB), while another stays on the RGB and the bottom of the AGB about
1 magnitude brighter than the HB of the cluster.

The observational studes of red giant variables gains are advanced
by the variable star survey programs and gravitational lensing
survey projects started in recent years, such as the {\it Long-Term
Photometry of Variables} (Jorissen et al. 1997), the {\it All Sky
Automated Survey} (Pojmanski 2002) and the {\it Northern Sky Survey}
(Wozniak et al. 2004) etc., especially,  MACHO (Alcock et al. 2000,
Wood et al. 1999, Schultheis 2004), OGLE (Kiss \& Bedding 2003, 2004, Wrag et al.
2004, Soszy\'nsky et al. 2004), EROS (Cioni et al. 2001) and MOA
(Bond et al. 2001, Noda et al. 2002) have accumulated a huge time
sequenced stellar photometric data set that is suitaable for searching
of variable stars. Wood (2000) found 5 distinct period-luminosity
sequences (A, B, C, D \& E, as he defined) in the
K--$\log P$ diagram based on his analysis of the MACHO near infrared
J \& K broadband photometry data in the half square degree field of
the LMC bar from. In that, the sequence C was the Mira variables
identified as the fundamental radial mode. The stars on the sequence B
wooud be known as semi-regular (SR) variables and A are of very low
amplitude as small amplitude red variables (SARVs). The  sequence B
and A correspond to pulsation in first-third overtines (Wood, 2000).
Such result is later confirmed independently by OGLE observations (Kiss
\& Bedding 2003, 2004). It is further pointed out by Soszy\'nski et
al. (2004) that the OGLE small amplitude variable red giants
(OSARGs) and the long period variables (LPVs) are two distinct types
of pulsating red variables.

The excitation mechanism of the pulsating red variables are far from fully
understood, the principal barrier is that there is not yet a perfect
time-dependent convection theory that can handle the coupling
between convection and oscillations. Most of the studies either
simply neglected the coupling between convection and oscillations,
such as Ostlie \& Cox (1986) and Gong et al. (1995), or
over-simplified it, such as Cox \& Ostlie (1991), Fox \& Wood
(1982). The most careful treatment of the coupling between
convection and oscillations comes from Balmforth (1992) and Xiong,
Deng \& Cheng (1998), with the former study applied Gough's (1977)
non-local time-dependent mixing-length theory of convection, while
the later one adopted Xiong's (1989a) time-dependent statistical
theory of convection. In our case, Xiong, Deng \& Cheng (1998)
studied the oscillations of Mira variables. The results show that
for the luminous red giants, the oscillations tend to happen in the
fundamental and low overtones whose amplitude growth rate is very
large, i.e. the amplitude gets doubled within a few, or sometimes
even within a single period. All the overtones above 4th order are
stable. For intermediate and low luminosity red giants, however, the
fundamental mode and the 1st overtone are stable while being
unstable at modes above the 2nd overtone (Xiong \& Deng 2001a).
These SARVs have rather small amplitude growth rates, being almost
neutrally stable, therefore special care is needed to handle
convection in this situation. The goal of the present work is to
deliver a thorough study of pulsation stability in stars
with intermediate- and low-luminosity on
the RGB and AGB. The second section describes the refinement and
specification of the convection theory for the current work, and the
third section presents the results of linear non-adiabatic
computations. The excitation mechanism for the red variables is
discussed in section 4. The conclusions of this work and discussions
are given in the last section.

\section{The fundamental equations}

Stars near the Hayashi limit (RGB and AGB) all have very deep
convective envelope. Instead of the $\kappa$-mechanism operating
in the hot and warm stars, the coupling between
convection and oscillations will be the primary mechanism of
excitation and damping for these cold stars. The main difficulty
prevent us from understanding the oscillations of such stars has been
the lack of a perfect convection theory. Being distinct from the
pulsational properties in the even cooler and brighter long period
variables, the pulsation amplitudes of the intermediate and low
luminosity red giants should be very small if any unstable p-mode
oscillations can be excited. This fact actually requires very
careful treatment of the coupling between convection and
oscillations, therefore makes the problem especially challenging.

Convection transfers
energy and momentum inside stars, and causes mixing of matters.
A star,as a thermodynamic system, can also be regarded as a
heating (or cooling) machine when considering its thermodynamic
behavior. Based on its nature and mechanism of working, the coupling
between convection and oscillations can be divided into two types as
the following:

\begin{enumerate}

\item The thermodynamic coupling between convection and oscillations,
whose mechanism is to adjust the temperature and pressure of stellar
material through its energy transfer function, therefore affects the
pulsational stability of the star. Its general effect is damping.
The thermodynamic coupling is the major reason for the existence of
the red edge of the Cepheid instability strip (including RR Lyrae
and $\delta$ Scuti) in the HRD (Xiong, Cheng \& Deng 1998, Xiong \&
Deng 2001a).

\item The dynamic coupling between convection and oscillations, whose
mechanism is to trigger momentum transfer within a star, i.e. to
affect the stability through turbulent pressure and turbulent
viscosity. Turbulent viscosity is always a damping, while turbulent
pressure is generally a excitation for oscillations. Such coupling
plays a key role in exciting the red
pulsating variable stars (Xiong, Deng \& Cheng 1998).

\end{enumerate}

The relative contributions of convective energy transfer, turbulent
pressure and turbulent viscosity to the instability of low
temperature srars change with temperature, luminosity and
pulsating mode of the stars, and this fact can be used to explains the complex
behaviors of pulsation instability among these stars having
extended convective envelopes. It is therefore quite easy to
understand such questions as follows: Why does the Ceipheid instability strip
have a red edge? Why do the long period variables exist outside the
Cepheid instability strip? What makes the luminous red variable
stars to have higher amplitudes and to pulsate at the fundamental and lower order
overtones than other stars, and what is the reason for the
decreasing of amplitudes in the red variables with decreasing
luminosity and eventually becoming pulsationally stable?

The main challenge in handling the coupling between convection and
oscillations is the requirement of a time-dependent theory of
convection. Owing to the huge scale of stars, stellar convection is
generally fully developed turbulence which is still not yet
understood satisfactorily. The main difficukty is due to the
nonlinearity and the non-locality of hydrodynamics. Compared to the
thermodynamic coupling between convection and oscillations, the
dynamical coupling is even more difficult to deal with, for the
anisotropic of stellar convection. The convection theory applied in
this work is a time-dependent one. Neglecting stellar rotation and
magnetic field, the full set of equations used in the calculations
of stellar convective envelope structure and radial oscillations is
the following,

\bq \frac{\partial r^3}{\partial M_r}=\frac{3}{4\pi
\bar{\rho}}\label{eq1}
\eq

\bqa \lefteqn{\frac{D\overline{u_r}}{DM_r}=-4\pi
r^2\frac{\partial}{\partial
M_r}\left(\bar{\rho}+\bar{\rho}\chi^2\right)}\nonumber \\
& &\mbox{} -{1\over r}\frac{\partial}{\partial M_r}\left(4\pi
r^3\bar{\rho}\Pi^{11}\right)-\frac{GM_r}{r^2},\label{eq2} \eqa

\bqa \lefteqn{C_P\frac{D\bar{T}}{Dt}+{3\over
2}\frac{D\chi^2}{Dt}-{B\over{\bar{\rho}}}\frac{D\bar{P}}{Dt}
-\frac{\chi^2}{\bar{\rho}}\frac{D\bar{\rho}}{DT}=}\nonumber \\
& &\mbox{}-\frac{\partial}{\partial M_r}\left(L_r+L_c+L_t\right),
\label{eq3}\eqa

\bq L_r=-\frac{16\pi^2 acr^4}{3\kappa}\frac{\partial
{\bar{T}}^4}{\partial M_r},\label{eq4} \eq

\bqa\lefteqn{{3\over
2}\frac{D\chi^2}{Dt}-\frac{\chi^2}{\bar{\rho}}\frac{D\bar{\rho}}{Dt}
+4\pi\bar{\rho}r^3\Pi^{11}\frac{\partial}{\partial
M_r}\left(\frac{\bar{u_r}}{r}\right)}\nonumber \\
& &\mbox -BV\left(\frac{D\bar{u_r}}{Dt}+\frac{GM_r}{r^2}\right)
+\frac{\partial}{\partial M_r}\left(4\pi
r^2\bar{\rho}\overline{u'_rw'_iw'^i/2}\right)\nonumber \\
& &\mbox
=-1.56\frac{GM_r\bar{\rho}\chi^3}{c_1r^2\bar{P}}\label{eq5}\eqa

\bqa
\lefteqn{\frac{DZ}{Dt}+2Z\left\{\left[1-B+\left(\frac{\partial\ln
C_P}{\partial\ln
T}\right)_T\right]\frac{D\ln\bar{T}}{Dt}\right. }\nonumber \\
  & &\mbox +\left.\left[\left(B-1\right)\nabla_{ad}
  +\left(\frac{\partial\ln C_P}{\partial\ln P}\right)_T\right]
  \frac{D\ln\bar{P}}{Dt}\right\},\nonumber\\
  & &\mbox +8\pi
r^2\bar{\rho}V\left(\frac{\partial\ln\bar{T}}{\partial
M_r}-\nabla_{ad}\frac{\partial\ln\bar{P}}{\partial
M_r}\right)\nonumber \\
  & &\mbox
+{1\over{\bar{\rho}\bar{C_P}^2}}\frac{\partial}{\partial
M_r}\left[4\pi
r^2\bar{\rho}^2\bar{C_P}^2\overline{u'_r\left(\frac{T'}{\bar{T}}\right)^2}\right]
\nonumber \\
  & &\mbox
=-1.56\frac{GM_r\bar{\rho}}{c_1r^2\bar{P}}\left(\chi+\chi_c\right)Z,\label{eq6}\eqa

\bqa \lefteqn{\frac{DV}{Dt}+V\left\{4\pi
r^2\bar{\rho}\frac{\partial\bar{u_r}}{\partial
M_r}+\left[1-B+\left(\frac{\partial\ln C_P}{\partial\ln
T}\right)_P\right]\frac{D\ln\bar{T}}{Dt}\right.}\nonumber \\
  & &\mbox +\left.\left[\left(B-1\right)\nabla_{ad}
  +\left(\frac{\partial\ln C_P}{\partial\ln P}\right)_T\right]
  \frac{D\ln\bar{P}}{Dt}\right\}\nonumber \\
  & &\mbox -BZ\left(\frac{D\bar{u_r}}{Dt}+\frac{GM_r}{r^2}\right)
  +4\pi r^2\bar{\rho}\chi^2\left(\frac{\partial\ln\bar{T}}{\partial M_r}
  -\nabla_{ad}\frac{\partial\ln\bar{P}}{\partial
  M_r}\right)\nonumber \\
  & &\mbox +\frac{1}{C_P}\frac{\partial}{\partial M_r}
  \left(4\pi r^2\bar{\rho}C_P
  \overline{u'^2_r\frac{T'}{\bar{T}}}\right)
  \nonumber \\
  & &\mbox =-0.78\frac{GM_r\bar{\rho}}{c_1r^2\bar{P}}
  \left(3\chi+\chi_c\right)V,\label{eq7}\eqa

\bqa \lefteqn{\frac{\Pi^{11}}{Dt}+{4\over 3}\pi r^3\bar{\rho}\chi^2
\frac{\partial}{\partial
M_r}\left(\frac{\bar{u_r}}{r}\right)}\nonumber \\
  & &\mbox +{4\over 3}\Pi^{11}\left[4\pi r^3\bar{\rho}\frac{\partial}{\partial M_r}
  \left(\frac{\bar{u_r}}{r}\right)+\frac{3\bar{u_r}}{2r}\right]\nonumber
  \\
  & &\mbox -{4\over 3}BV\left(\frac{D\bar{u_r}}{Dt}+\frac{GM_r}{r^2}\right)
  +\frac{\partial}{\partial M_r}\left(4\pi r^2\bar{\rho}\overline{u'_r\Pi'^{11}}\right)
  \nonumber \\
  & &\mbox =-{4\over
  3}\frac{0.78\left(1+c_3\right)GM_r\bar{\rho}}{c_1r^2\bar{P}}\chi \Pi^{11},
  \label{eq8}\eqa

where r is the radius, $M_r$ is the mass within the sphere of radius
r, $u_r$ is the radial component of the velocity, $\rho$, T and P
are respectively the regular notations for density, temperature and
pressure (including radiative), $\chi$ is the opacity, $C_P$ is the
specific heat at constant pressure,
$B=-\left(\partial\ln\rho/\partial\ln T\right)_P$ is the expansion
coefficient of gas. Our formulism of stellar convection is a
statistical theory of correlation functions based on hydrodynamics
and the theory of turbulence, $\chi^2$ and $\Pi^{ij}$ are respectively
the isotropic and the anisotropic components of the auto-correlation
of turbulent velocity $w'^i$, therefore we have,

\bq \overline{w'^iw'^j}=g^{ij}\chi^2+\Pi^{ij},\label{eq9}\eq

$Z$ and $V$ are respectively the auto-correlation of turbulent
temperature fluctuation $T'/\bar{T}$ and the cross-correlation of
radial turbulent velocity and temperature, defined as,

\bq Z=\overline{\left(\frac{T'}{\bar{T}}\right)^2},\label{eq10}\eq
\bq V=\overline{w'_r\frac{T'}{\bar{T}}}.\label{eq11}\eq

In our equations, all the quantities with a prim are the turbulent
fluctuations, while those with a bar are statistical averages
of the corresponding quantities.

Eqs.~(\ref{eq1})--(\ref{eq8}) are derived based the equations of
hydrodynamics and the theory of turbulence, where
Eqs.~(\ref{eq1})--(\ref{eq4}) are the normal equations for stellar
structure, being different from the traditional expressions only by
having the turbulent Renold's stress (turbulent pressure and
viscosity) included in eq.~(\ref{eq2}) and having both the turbulent
thermal convective flux $L_c=4\pi r^2\bar{\rho}C_P\bar{T}V$ and the
turbulent kinetic energy flux $L_t=4\pi
r^2\bar{\rho}\overline{u'_rw'_iw'^i/2}$ included in eq.~(\ref{eq3}).
As a result of this consideration, the correlations of turbulent
convection also show up eqs.~(\ref{eq2}) and (\ref{eq3}).
Eqs.~(\ref{eq5})--(\ref{eq8}) are a set of dynamical equations of
turbulent convection under non-local theoretical frame.
Owing to the fact that the equations of stellar
structure include the convective  quantities and
the stellar structural quantities, $\bar{\rho}$,
$\bar{P}$ and $\bar{T}$ (and $\bar{u_r}$ as well) enter the
dynamical equations of turbulent correlations of
eqs.~(\ref{eq5})--(\ref{eq8}), it makes these equations coupled with
each other, therefore eqs.~(\ref{eq1})--(\ref{eq8})
formed a complete set of equations for the structure and
oscillations of stellar envelope at spherical symmetry condition (of
course without rotation and magnetic field, as we assumed at the
beginning). In the frame work of the local theory, stellar
convection is handled with a set of algebraic equations, while under
non-local expressions it is dealt with by a set a differential (or
integral) equations. In nature, all fluid motions are in fact
non-local, the local convection approximation is good only at a
limited local region (such as the interior zones far away from the
boundaries of a convective zone). The local theory has no chance to
be right when considering a whole convective region as an entity.
This is especially true near the boundaries and the overshooting
zones where the local approximation should be completely lifted, and
in this context, one has to use a non-local theory. Our non-local
theory of stellar convection is a statistical theory of turbulent
correlations derived from hydrodynamics and the theory of
turbulence, its physical picture is very clear.
Eqs.~(\ref{eq5})--(\ref{eq8}) describe a statistical equilibrium of
turbulent correlations. Turbulence gains energy from buoyant force
and the average of motion of fluid, this process happens in the low
wave number region of the turbulent spectrum. Due to the
non-linearity and interactions of fluid motion, turbulent energy is
cascaded towards higher wave numbers from lower wave numbers,
and is turned into thermal
energy by molecular viscosity and is eventually completely
dissipated. The right hand side terms including the convection
parameter $c_1$ in eqs.~(\ref{eq5})--(\ref{eq8}) represent here the
turbulent viscous dissipation. following the turbulence
theory the dissipation rate of turbulent kinetic energy
$\bar{\rho}\epsilon$ can be expreessed as follows (Hinze 1975),

\bq\epsilon=2\eta_ek_e\chi^3,\label{eq12}\eq

where $k_e=\sqrt{3}k_{e1}=\sqrt{3}/l_{e1}$ is the wave number of the
energy-containing eddies. It is assumed in this work that $l_{e1}$
is proportional to the local pressure scale height, i.e.,

\bq l_{e1}=c_1H_P=c_1\frac{r^2P}{GM_r\bar{\rho}}.\label{eq13}\eq

The third order correlations in eqs.~(\ref{eq5})--(\ref{eq8})
correspond to the non-local convective diffusion. The third order
correlation term in eq.~(\ref{eq5}), for example, is just the
turbelent kinetic energy flux. A gradient type of diffusion
approximation for the third order correlations is adopted in our
theory of non-local convection. Taking the third order correlation
for turbulent velocity as a instance, we have,

\bq\L_t=4\pi r^2\bar{\rho}\overline{u'_rw'_iw'^i/2}=-6\pi
r^2\bar{\rho}\chi\Lambda\frac{\partial \chi^2}{\partial
r},\label{eq14}\eq

where $\Lambda$ is the diffusion length assumed to be proportional
the the local pressure scale height,

\bq\Lambda=\frac{\sqrt{3}}{4}c_2H_P={\sqrt{3}\over
4}c_2\frac{r^2\bar{P}}{GM_r\bar{\rho}},\label{eq15}\eq

where $c_2$ is another convective parameter that is linked to the
non-local turbulent diffusion process.

Putting eq.~(\ref{eq15}) into eq.~(\ref{eq14}), $L_t$ can be
rewritten as,

\bq L_t=-{3\over 2}Q\frac{\partial\chi^2}{\partial
M_r},\label{eq16}\eq

where,

\bq
Q=\frac{4\sqrt{3}\pi^2c_2r^6\bar{\rho}\bar{P}\chi}{GM_r}.\label{eq17}\eq

Following similar operations, the remaining third order correlations
in eqs.~(\ref{eq6})--(\ref{eq8}) can be given as,

\bq 4\pi
r^2\bar{\rho}^2C_P^2\overline{u'_r\left(\frac{T'}{\bar{T}}\right)^2}
=-\bar{\rho}C_P^2Q\frac{\partial Z}{\partial M},\label{eq18}\eq

\bq 4\pi
r^2\bar{\rho}C_P\overline{{u'^2}_r\left(\frac{T'}{\bar{T}}\right)}
=-C_PQ\frac{\partial V}{\partial M_r},\label{eq19}\eq

\bq 4\pi r^2\bar{\rho}\overline{u'_r\Pi '^{11}}=-Q\frac{\partial
\Pi^{11}}{\partial M_r}. \label{eq20}\eq

As described above, in the static station convection gains energy
from buoyant force, therefore the initial convective motion is in
radial direction. However, due to the continuity of fluid motion and
the non-liner interactions among the turbulent eddies, a part of the
initial turbulent kinetic energy will be converted into the
horizontal directions. It is known from the theory of turbulence
that the correlation of turbulent pressure and velocity gradient
tends to make turbulence isotropic (Rotta 1951). The convection
parameter $c_3$ in eq.~(\ref{eq8}) is introduced to describe the
tendency of restoring the isotropy. The larger $c_3$, the stronger
the restoring force and the turbulence is therefore more isotropic.
Therefore, the convection parameter $c_3$ can be regarded as the
parameter for the degree of isotropy of turbulence. In the
convectively unstable zone, the ratio of the radial component of
turbulence ($w'_r$) and the horizontal one ($w'_h$) is
$\overline{w'^2_r}/\overline{w'^2_h}\approx\left(3+c_3\right)/2c_3$
(Deng \& Xiong 2006).

Eqs.~(\ref{eq1})--(\ref{eq8}) are a completed set of dynamical
equations of stellar envelope structure, which can be  used to
construct the equilibrium envelope model and to calculate the radial
oscillations. When setting all the terms involving time derivatives
and velocity $\bar{u}_r$ in these equations to be zero, we will get
the equations for equilibrium model. When a star does
small amplitudes oscillations around the equilibrium, all the
quantities can be made to be the sum of the equilibrium values and
the perturbation (oscillations), then a Tyler expansion around the
equilibrium can be applied. By conserving only the linear terms and
neglecting all terms over the first order, we will get the equations
of linear non-adiabatic oscillations. Details on the equations for
equilibrium model and linear pulsation, and the method for
solving these equations are not given here, for those who are
interested in that, our previous work is referred (Xiong \& Deng
2001b, Xiong, Deng \& Cheng 1998). It needs to be noticed that we
assumed quasi-isotropy for turbulent convection in our previous
work. In the current, turbulent convection is more carefully
handled, with anisotropy of turbulence included, i.e. a equation
designated to the anisotropy of turbulent convection (\ref{eq8}) is
now added to the equations for both the equilibrium model and
oscillations.

\section{Numerical results of linear non-adiabatic oscillations for red giants}

Making use of the equations introduced in the previous section,
the linear non-adiabatic oscillations of 10 sequences of evolutionary models
with masses $M=1,1.2,1.4,1.6,1.8,2.0,2.25,2.5 \& 2.75M_\odot$ are calculated.
The chemical composition of all these models is solar, $X=0.70$ and
$Z=0.02$. The evolutionary tracks are made using Padova code
(Bressan, et al. 1993). For selected stellar models along the tracks
with given luminosity and effective temperature, the corresponding
non-local convective models of equilibrium envelope are constructed, then
linear non-adiabatic oscillations for the fundamental through
11th overtones are further calculated. Special care is taken when
handling thermodynamical and dynamical coupling between convection
and oscillation as describe in detail in the previous section.
Table~\ref{table1} gives, as an example, the amplitude growth rate $r=-2\pi
\omega_i/\omega_r$ of linear non-adiabatic oscillations for one of
the evolutionary model with mass $M=1.0M_\odot$, where $\omega_i$ and
$\omega_r$ are respectively the imaginary and the real component of
the complex frequency of oscillations $\omega=i\omega_i+\omega_r$.

\begin{table*}
\begin{minipage}{128mm}
\caption{Amplitude growth rates for a set of evolutionary models of
star with 1M$_\odot$}\label{table1}
\begin{tabular}{@{}crrrrrrrrrrr}
\hline No & Log Te & Log L & RATE0 &  RATE1 & RATE2 & RATE3 & RATE4
& RATE5 & RATE6 & RATE7 & RATE8\\ \hline
1&3.7535&-0.1591&-0.15E-8&-0.16E-7&-0.82E-7&-0.31E-6&-0.93E-6&
-0.22E-5&-0.48E-5&-0.97E-5&-0.19E-4\\2&3.7460&-0.2022&-0.12E-8&
-0.12E-7&-0.64E-7&-0.24E-6&-0.73E-6&-0.18E-5&-0.39E-5&
-0.79E-5&-0.16E-4\\3&3.7365&-0.2256&-0.81E-9&-0.10E-7&-0.53E-7&
-0.20E-6&-0.62E-6&-0.15E-5&-0.33E-5&-0.69E-5&-0.13E-4\\4&
3.7264&-0.2407&-0.13E-8&-0.82E-8&-0.46E-7&-0.18E-6&-0.54E-6&
-0.14E-5&-0.30E-5&-0.62E-5&-0.12E-4\\5&3.7550& 0.3154&-0.17E-7&
-0.19E-6&-0.93E-6&-0.33E-5&-0.89E-5&-0.19E-4&-0.40E-4&
-0.75E-4&-0.93E-4\\6&3.7452&0.1800&-0.57E-8&-0.72E-7&-0.39E-6&
-0.14E-5&-0.41E-5&-0.94E-5&-0.20E-4&-0.42E-4&-0.64E-4\\7&
3.7500&0.0048&-0.33E-8&-0.32E-7&-0.17E-6&-0.63E-6&-0.19E-5&
-0.44E-5&-0.92E-5&-0.19E-4&-0.36E-4\\8&3.7412& 0.3459&-0.12E-7&
-0.17E-6&-0.93E-6&-0.34E-5&-0.90E-5&-0.20E-4&-0.45E-4&
-0.79E-4&-0.80E-4\\9&3.7353&0.3515&-0.10E-7&-0.16E-6&-0.89E-6&
-0.32E-5&-0.87E-5&-0.20E-4&-0.45E-4&-0.77E-4&-0.74E-4\\10&
3.7307&0.3554&-0.89E-8&-0.15E-6&-0.86E-6&-0.32E-5&-0.85E-5&
-0.20E-4&-0.46E-4&-0.78E-4&-0.73E-4\\11&3.7268&0.3548&-0.48E-8&
-0.14E-6&-0.81E-6&-0.30E-5&-0.81E-5&-0.19E-4&
-0.44E-4&-0.72E-4&-0.63E-4\\12&3.7217&0.3520&-0.56E-8&
-0.12E-6&-0.75E-6&-0.28E-5&-0.76E-5&-0.18E-4&-0.42E-4&
-0.68E-4&-0.58E-4\\13&3.7156&0.3514&-0.47E-8&-0.11E-6&
-0.69E-6&-0.26E-5&-0.71E-5&-0.18E-4&-0.41E-4&-0.64E-4&
-0.52E-4\\14&3.7107&0.3492&-0.48E-8&-0.98E-7&-0.63E-6&
-0.24E-5&-0.67E-5&-0.17E-4&-0.39E-4&-0.60E-4&-0.49E-4\\15&
3.7067&0.3502&-0.43E-8&-0.99E-7&-0.63E-6&-0.24E-5&-0.65E-5&
-0.16E-4&-0.35E-4&-0.48E-4&-0.24E-4\\16&3.7009&0.3512&-0.33E-8&
-0.99E-7&-0.64E-6&-0.24E-5&-0.64E-5&-0.15E-4&
-0.32E-4&-0.36E-4&0.42E-5\\17&3.6962&0.3590&-0.31E-8&-0.10E-6&
-0.66E-6&-0.24E-5&-0.64E-5&-0.15E-4&-0.27E-4&-0.19E-4&0.42E-4
\\ 18&3.6914&0.3747&-0.39E-8&-0.11E-6&-0.71E-6&-0.26E-5&
-0.66E-5&-0.15E-4&-0.23E-4&-0.23E-5&0.81E-4\\19&3.6869&0.4042&
-0.41E-8&-0.13E-6&-0.83E-6&-0.30E-5&-0.74E-5&-0.16E-4&
-0.18E-4&0.20E-4&0.14E-3\\20&3.6800&0.4915&-0.65E-8&-0.20E-6&
-0.13E-5&-0.44E-5&-0.11E-4&-0.18E-4&0.46E-5&0.10E-3&0.36E-3\\
21&3.6750&0.6207&-0.98E-8&-0.40E-6&-0.25E-5&-0.82E-5&-0.18E-4&
-0.17E-4&0.64E-4&0.31E-3&0.84E-3\\22&3.6700&0.7916&-0.22E-7&
-0.10E-5&-0.60E-5&-0.18E-4&-0.31E-4&0.37E-4&0.31E-3&0.10E-2&
0.19E-2\\23&3.6650&0.9496&-0.52E-7&-0.24E-5&-0.14E-4&-0.37E-4&
-0.28E-4&0.21E-3&0.93E-3&0.22E-2&0.34E-2\\24&3.6550&1.1935&
-0.24E-6&-0.93E-5&-0.43E-4&-0.62E-4&0.28E-3&0.14E-2&0.36E-2&
0.53E-2&0.77E-2\\25&3.6500&1.3335&-0.58E-6&-0.20E-4&-0.75E-4&
0.14E-4&0.84E-3&0.31E-2&0.57E-2&0.81E-2&0.11E-1\\26&3.6450&
1.4252&-0.10E-5&-0.31E-4&-0.10E-3&0.18E-3&0.15E-2&0.48E-2&
0.72E-2&0.11E-1&0.13E-1\\27&3.6400&1.5126&-0.18E-5&-0.47E-4&
-0.11E-3&0.54E-3&0.27E-2&0.68E-2&0.95E-2&0.14E-1&0.15E-1\\28&
3.6350&1.5835&-0.34E-5&-0.70E-4&-0.16E-4&0.13E-2&0.51E-2&
0.90E-2&0.13E-1&0.16E-1&0.17E-1\\29&3.6300&1.6545&-0.41E-5&
-0.87E-4&-0.31E-5&0.15E-2&0.57E-2&0.97E-2&0.14E-1&0.17E-1&
0.17E-1\\30&3.6275&1.6789&-0.31E-5&-0.63E-4&0.36E-3&0.29E-2&
0.90E-2&0.13E-1&0.18E-1&0.18E-1&0.15E-1\\31&3.6217&1.7407&
-0.48E-5&-0.92E-4&0.36E-3&0.29E-2&0.90E-2&0.13E-1&0.19E-1&
0.21E-1&0.21E-1\\32&3.6150&1.8503&-0.12E-4&-0.16E-3&0.81E-3&
0.45E-2&0.11E-1&0.15E-1&0.20E-1&0.19E-1&0.18E-1\\33&3.6101&
1.9072&-0.15E-4&-0.95E-4&0.20E-2&0.73E-2&0.15E-1&0.19E-1&
0.22E-1&0.20E-1&0.18E-1\\34&3.6065&1.9400&-0.12E-4&0.34E-5&
0.29E-2&0.97E-2&0.19E-1&0.25E-1&0.30E-1&0.27E-1&0.22E-1\\35&
3.6010&2.0037&-0.12E-4&0.18E-3&0.45E-2&0.13E-1&0.22E-1&
0.30E-1&0.32E-1&0.28E-1&0.18E-1\\36&3.5961&2.0656&-0.22E-4&
0.47E-4&0.37E-2&0.12E-1&0.18E-1&0.27E-1&0.27E-1&0.22E-1&
0.74E-2\\37&3.5925&2.1176&-0.18E-4&0.84E-3&0.67E-2&0.17E-1&
0.21E-1&0.26E-1&0.22E-1&0.17E-1&0.68E-2\\38&3.5900&2.1346&
-0.24E-4&0.35E-3&0.52E-2&0.15E-1&0.20E-1&0.29E-1&0.26E-1&
0.18E-1&-0.21E-2\\39&3.5813&2.2610&0.10E-3&0.42E-2&0.18E-1&
0.40E-1&0.43E-1&0.51E-1&0.35E-1&0.78E-2&-0.19E-1\\40&3.5534&
2.5591&0.13E-2&0.31E-1&0.62E-1&0.78E-1&0.75E-1&0.25E-1&
-0.22E-1&-0.23E-1&-0.11E-1\\
\hline
\end{tabular}
\end{minipage}
\end{table*}

The equation of state adopted in this work is a simplified MHD
expression. The main simplification made here is that a helium atom
is still considered as a quasi-hydrogen atom, which makes the
calculation for the energy levels of helium atomic far easier. For
the first ionized helium atoms or the high level excitation states
of neutral helium atoms, this is no doubt a fairly good treatment.
This approximation turns out to have large deviation from the real
situation only for the low excitation states of neutral helium. It
is predictable that the quasi-hydrogen atom approximation only
brings into the equation of state and the relevant thermodynamic
quantities very limited affection, however, it has the very good
advantage that, instead of interpolated with the pre-computed
tables, the equation of state and the thermodynamic quantities can
be calculated on the mesh in real time. The radiative opacity used
in this work is a analytic formula to approach the OPAL tables
(Rogers \& Iglesias 1992) and the low temperatures opacity tables
(Alexander et al. 1994). This analytic expression gives a smooth
connection of OPAL table at intermediate and high temperature region
and that of Alexander et al. (1994) in low temperatures region. The
surface boundary for pulsation calculations are set at $\tau=0.01$,
and the bottom one is located at a temperature of $T\sim 5x10^6K$.

\subsection{Pulsation stability}

\begin{figure}
\includegraphics[width=84mm]{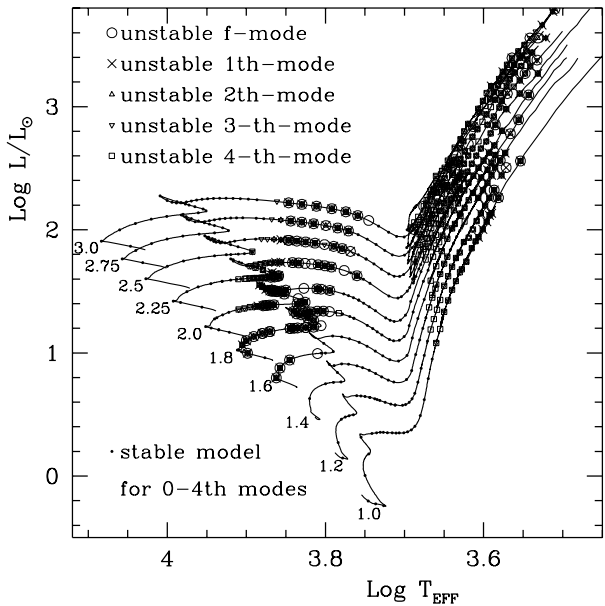}
\caption{The distribution of pulsationally stable and unstable stars
in the HRD. The tiny dots are pulsationally stable stars from the
fundamental up to 4th overtone, the open symbols of circles, plus,
triangles, inverse triangles and squares are respectively the
unstable stars in the fundamental up to the 4th
overtone.}\label{fig1}
\end{figure}

The distribution of the pulsationally stable and unstable models
in HRD from our calculations is shown in fig.~\ref{fig1},
where the tiny solid dots are stable models from the
fundamental through the 4th overtone ($r_i<0$, $i=1,4$); while the
open circles, crosses, triangles, inverse triangles and squares
are the unstable models respectively in the fundamental mode to
the 4th overtone. Two separate instability strips can be seen clearly
in the plot. The one of them in the middle of fig.~\ref{fig1} is the
well known $\delta$ Scuti instability strip which is not really the
topic of current work. As the $\delta$ Scuti strip is calculated
using the same scheme, showing it here assures the reliability of
our theory and the numerical scheme. The second one is the pulsating red giant
instability strip on the upper-right of the figure. All the models
between the two strips are all pulsationally stable at least for modes up to the
4th overtone. For RGB stars with low luminosity, the fundamental and
the first overtone are both stable. These stars are unstable only in
high order overtones. For high luminosity and lower effective
temperature stars (going up along the RGB and AGB), pulsation instability
shifts towards lower overtone and the fundamental mode.

It follows from table \ref{table1} and fig.~\ref{fig1} that all the
modes lower than the 5th order modes are stable for the
$M=1.0M_\odot$ RGB stars with $\log T_e\geq 3.66$. This agrees well
with the observations of Henry et al. (2000). For stars with higher
masses, the blue edge of the red giant instability strip shifts to
higher temperature and higher luminosity. For a $M=3M_\odot$ RGB
star, the blue edge of instability strip moved to $\log T_e\approx
3.68$.

It has to be bared in mind that the location of the blue edge of
the red giant instability strip depends not only on stellar mass,
but also on metallicity. For increasing (decreasing) metallicity,
the blue edge moves to red (blue). We did similar calculations for
intermediate or low luminosity stars in globular clusters
($Z=10^{-4}$) and we found that all RGB stars below the HB are
stable in the fundamental mode while oscillating in 2-4th overtones,
in that case the blue edge was shifted to $\log T_e=3.74$ (Xiong \&
Deng 2006a,b). The 2 small amplitude short period red variables K1098
and VZ1140 found in M15 and M3 (Yao \& Qin 1993, Yao, Zhang \& Qin
1993) are both located near the intersection of HB and RGB, and
their periods of the fundamental mode are both greater than 1 day
according to theoretical calculations. However, the observed periods
are respectively 0.5155 and 0.35 days, therefore suggesting that
they are very likely pulsating at 2nd to 4th orders overtone.
This agrees with the prediction of our work.

\subsection{Pulsation period and constant}

\begin{figure}
\includegraphics[width=84mm]{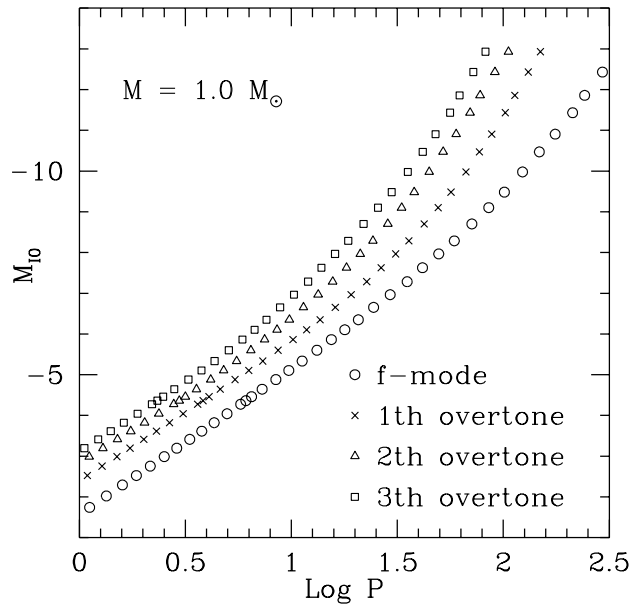}\caption{The period--luminosity relation for the
evolutionary sequence of a 1M$_\odot$ star. The horizontal axis is
the logarithmic period, and the vertical one is the the reddening
free Wesenheit index (Soszy\'nski et al. 2004). The circles, plus
signs, triangles and squares represent respectively the fundamental
up to 3rd overtone.}\label{fig2}
\end{figure}

The pulsation period can be derived from pulsation calculations.
It is then straightforward to build a theoretical period-luminosity
relation. However, to make such relation ready to be compared with
that of observations is not so easy. In fig.~\ref{fig2}, the
theoretical period-luminosity relation for a $M=1.0M_\odot$ RGB
model is given. The horizontal axis is the pulsation period, and the
vertical one is the reddening free Wesenheit index $W_I$
(Soszy\'nski et al 2004)

\[ W_I=I-1.55\left(V-I\right) \]

The temperature to color index conversion and the bolometric
correction is taken from Johnson (1966). It can be found that the
overall result of our calculations looks very similar to the
observational one made for the red giant variables in Magellanic
clouds (Soszy\'nski et al. 2004). Unfortunately, quantitatively
accurate comparison between our theoretical result and the observational
relation is not possible for now due to the following reasons,

\begin{enumerate}

\item The temperature of the RGB stars depends critically on the
metallicity, low temperature opacity and the mixing length parameter
$\alpha$. Larger metallicity and higher opacity, or smaller
mixing length parameter $\alpha$, will make the RGB move towards
lower effective temperature;

\item For low temperature RGB stars, both the effective temperature to
color index conversion and bolometric correction are the worst
determined, and this is especially true for AGB stars usually having
a circumstellar dust envelopes which complicated the problem;

\item The stars in the Magellanic clouds are not at all a group of
stars having the same mass, age and chemical composition. Instead,
they have all combinations of all those parameters.

\end{enumerate}

\begin{figure*}
\includegraphics[width=84mm]{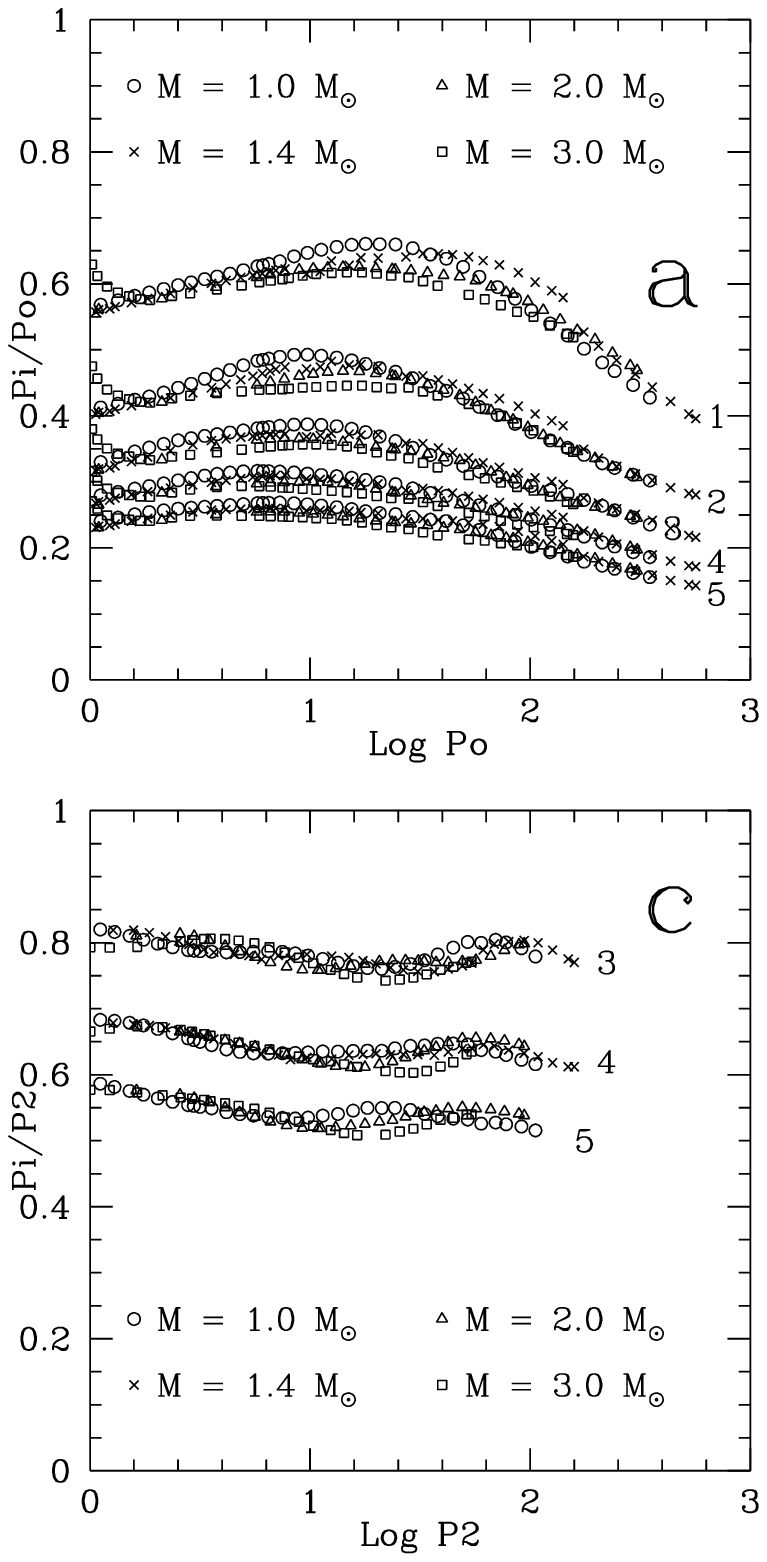}\includegraphics[width=84mm]{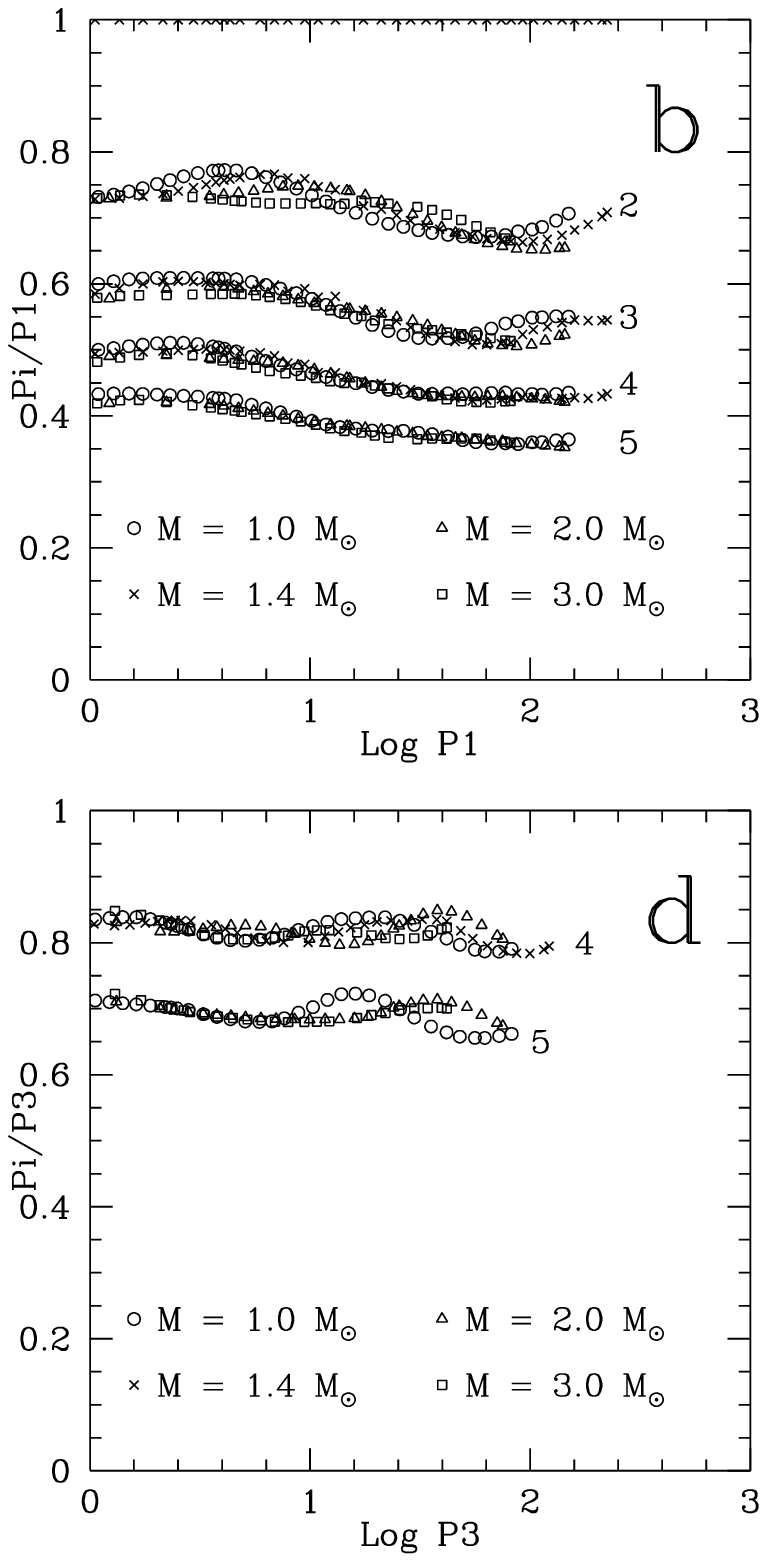}
\caption{The theoretical Peterson diagram for evolutionary model
sequence. Plotted are the period ratios $P_j/P_i$ ($i=0,1,2,3$;
$j=i-1,......5$) versus $\log P_i$. The 4 panels have different $i$
indices, a) $i=0$ the fundamental mode; b) $i=1$ the 1st overtone;
c) $i=2$ the 2nd overtone; and d) $i=3$ the 3rd overtone. The
circles, plus signs, triangles and squares are respectively the
sequences for M=1, 1.4, 2.0 and 3.0M$_\odot$.}\label{fig3}
\end{figure*}

In practice, the ratios between different overtones is a far better
predictable than the periods, because they are less affected by
stellar parameters. Figs.~\ref{fig3}a and \ref{fig3}d are the
theoretical Petersen diagrams from our calculations. A direct
comparison shows that our results  are very similar to the observations
for the OSARG in LMC (Soszy\'nski et al. 2004). Therefore we make a
further speculation. There are the 4 ridges of period-luminosity
relations in their fig.~7. We identify that the ridge (or sequence)
a$_1$ (b$_1$) is the 1st overtone, a$_2$ (b$_2$) is the 2nd
overtone, while a$_3$ (b$_3$) is a mixture of the 3rd and 4th
overtones, and a$_4$ and b$_4$ are the 5th overtone. In fact, P4 in
their paper should be the 5th instead of 4th overtone, because for
the 4th overtone, it is not possible to have ratios of P4/P1, P4/P2
and P4/P3 being as small as 0.39, 0.56 and 0.76 respectively.
However, if one considers that P3 in their paper is a mixture of 3rd
and 4th overtones, while P4 is the 5th one, then a perfect match
between the observations and our theoretical model.

\begin{figure*}
\includegraphics[width=84mm]{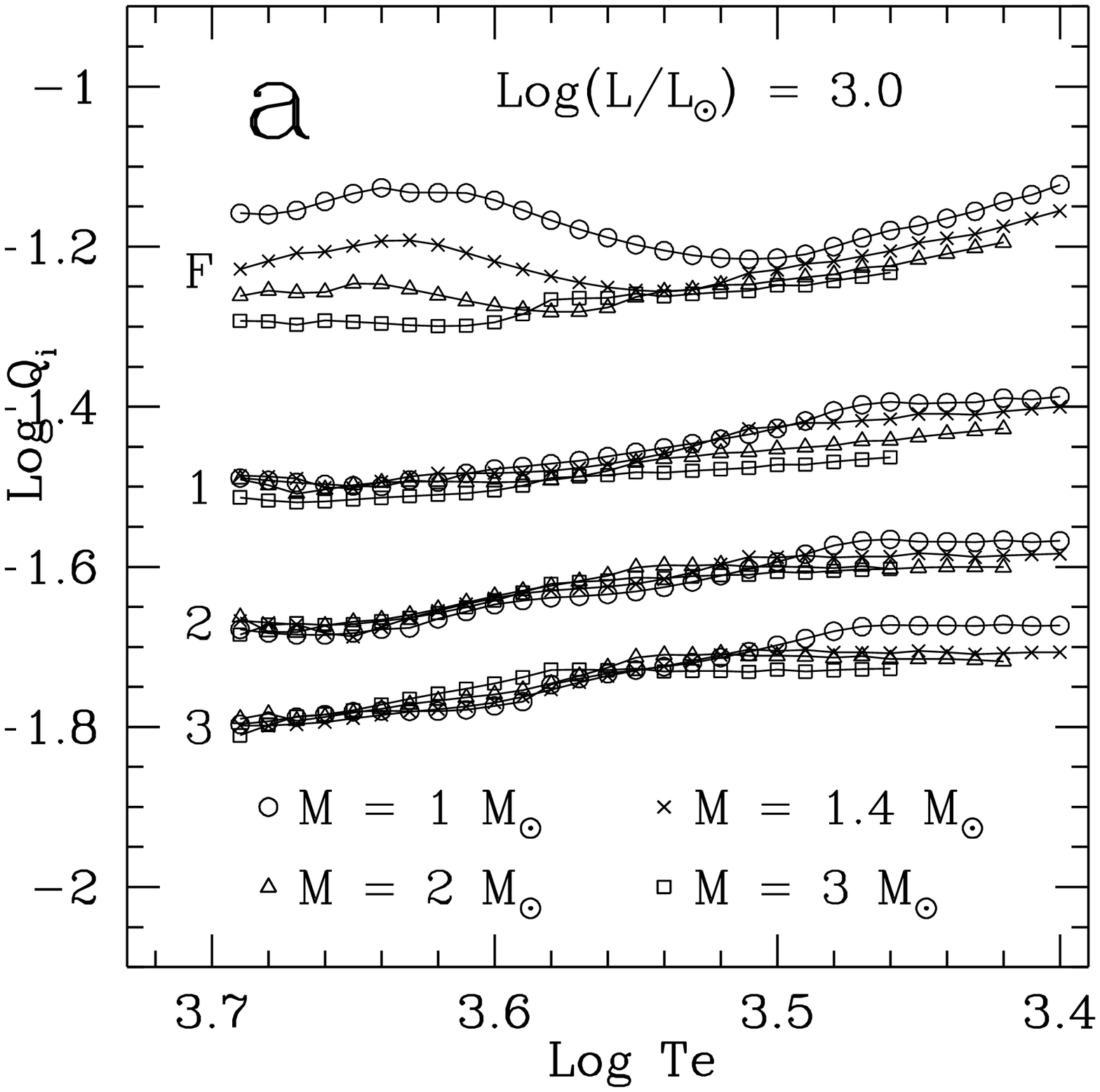}\includegraphics[width=84mm]{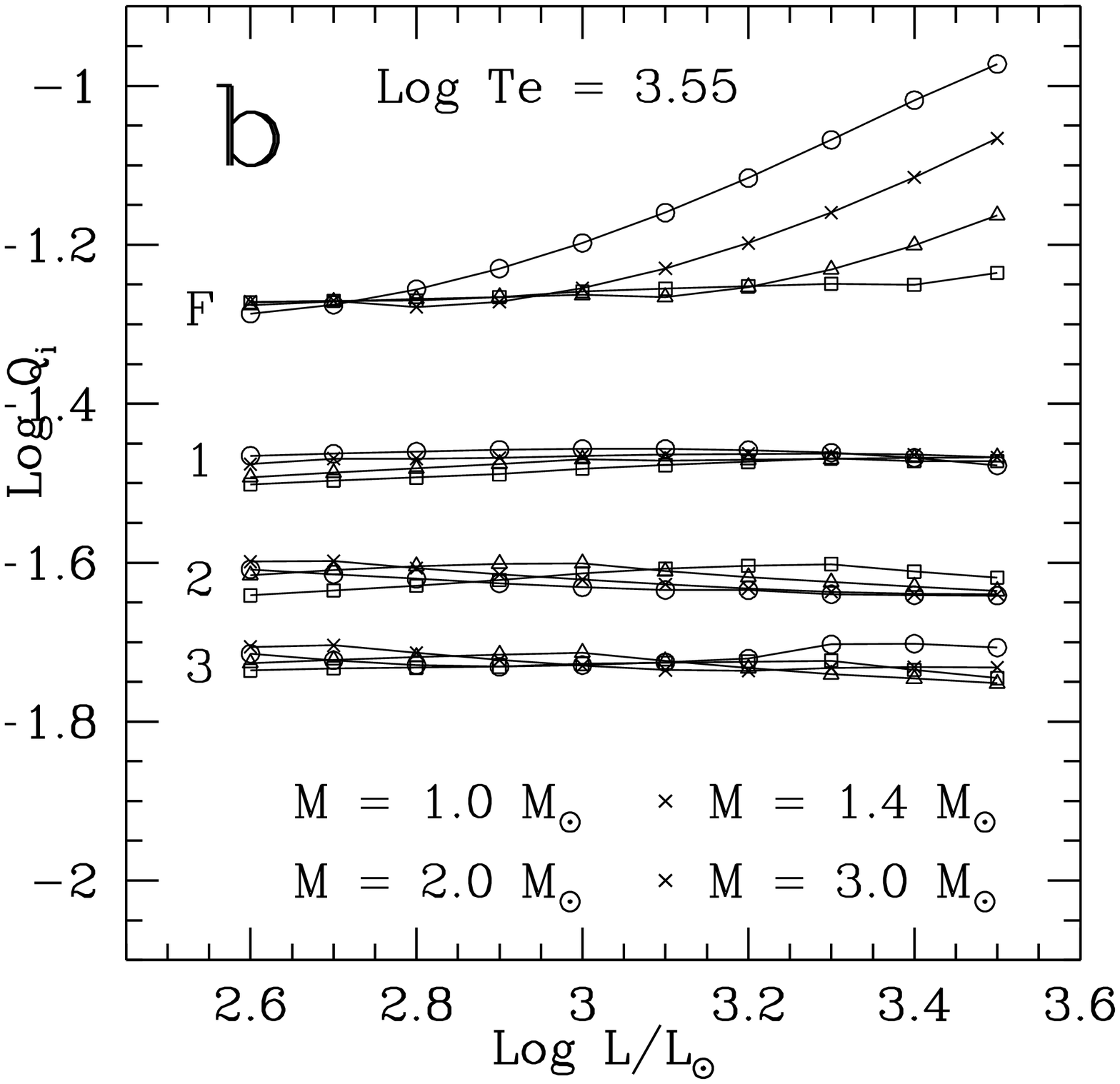}
\caption{a) The pulsation constants from the fundamental model to
the 3rd overtone mode as functions of effective temperature at a
given luminosity of $\log L/L_\odot=3.0$. b) The pulsation constants
as functions of effective luminosity at a given effective
temperature of $\log Te=3.55$. The circles, plus signs, triangles
and squares are respectively for M=1, 1.4, 2.0 and
3.0M$_\odot$.}\label{fig4}
\end{figure*}

Relative to pulsation period, the pulsation constant $Q$ changes
less, since it is only a slow varying function of stellar mass,
luminosity and effective temperature. Fig.~\ref{fig4}a shows the
pulsation constants $Q_i$ (i=0--3) of the fundamental mode through
the 3rd overtone as functions of effective temperature, calculated
for 4 sets of RGB models with masses of 1, 1.4, 2.0 and 3.0 $M_\odot$
respectively, and with the same luminosity $L=1000L_\odot$. From
fig.~\ref{fig4}a, the pulsation constants of 1st--3rd overtones
changes relatively smaller than the fundamental mode does. The pulsation
constants grow slowly with decrease of effective temperature for these
overtones. While for the fundamental mode, on the contrary, The
pulsation constants change largely with complex pattern, being
a non-monotonic function of effective temperature, but showing
a tendency of slight decrease with increase of mass.

Fig.~\ref{fig4}b shows the pulsation constant as a function of
luminosity for 4 sets of models of red giant models with the same
temperature $\log T_e=3.55$ but the different masses. Similarly, the
pulsation constants of the 1st--3rd overtones are almost unchanged,
while the fundamental mode shows a much larger variation, the
pulsation constants increase with increase of luminosity of stars.

To further explore how the pulsation constant changes with
stellar parameters such as mass, luminosity and effective
temperature, 24 sets of models of red giant are calculated
as a supplementary database for this aim. They consist of 600
models with diffirent masses ($M=1, 1.4, 2.0 and 3.0M_\odot$),
luminosites ($\log L/L_\odot=2.6-3.5$) and effective
temperature ($\log T_e=3.40-3.70$). we get the following
analytic fitting formula for pulsation constant,

\bqa\lefteqn{\log
Q=\sum^3_{n=0}a_nt^n+m\sum^3_{n=0}b_nt^n}\nonumber\\
 & &\mbox
 +\left[\sum^3_{n=0}c_nt^n+m\sum^3_{n=0}d_nt^n\right]l\nonumber\\
 & &\mbox
 +\left[\sum^3_{n=0}e_nt^n+m\sum^3_{n=0}f_nt^n\right]l^2,\label{eq21}
\eqa

where

\[ m=\log M/M_\odot, \]
\[ t=\log T_e -3.55, \]
\[ l=\log L/L_\odot -3.0.\]

\begin{figure*}
\includegraphics[width=84mm]{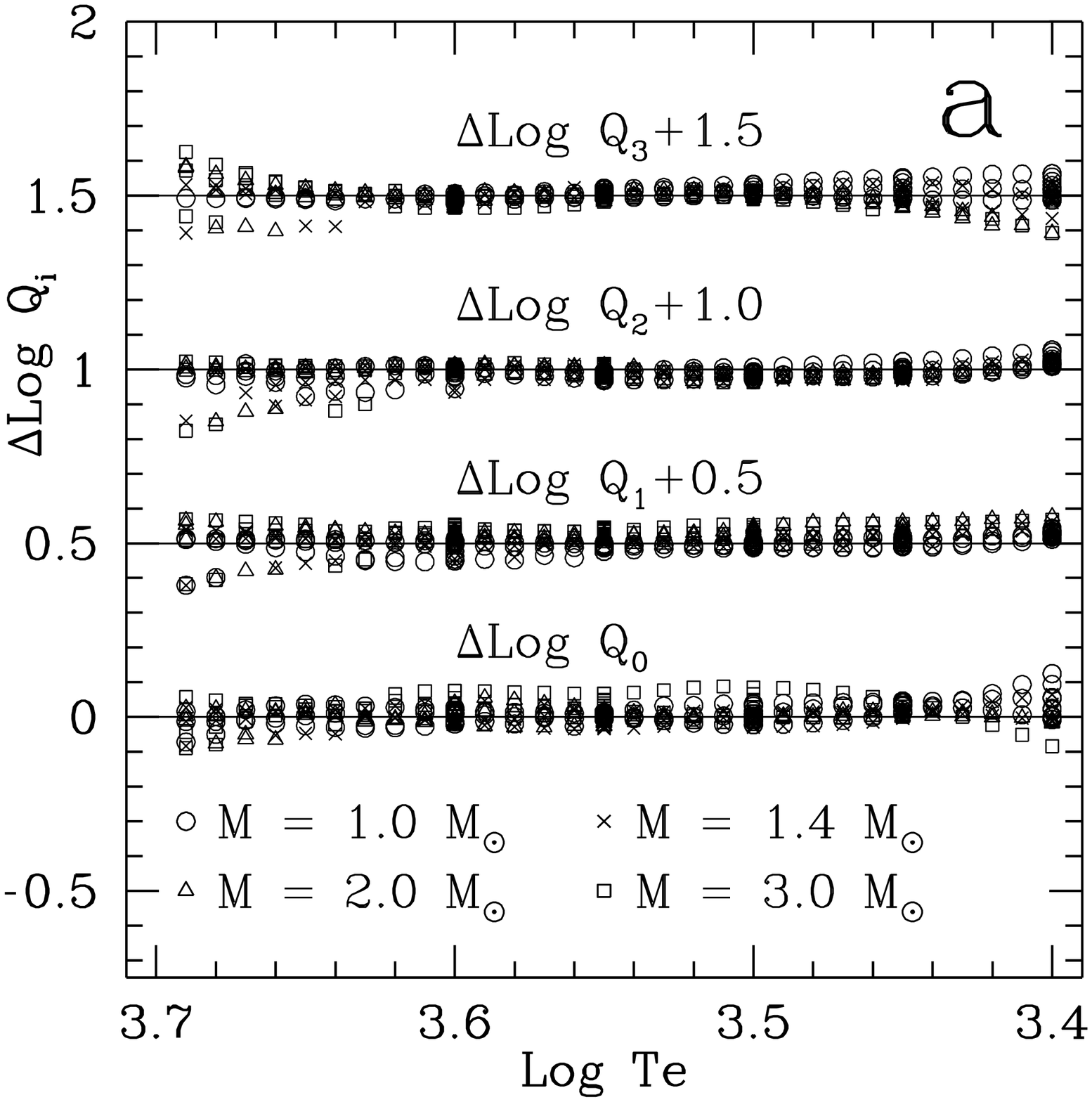}\includegraphics[width=84mm]{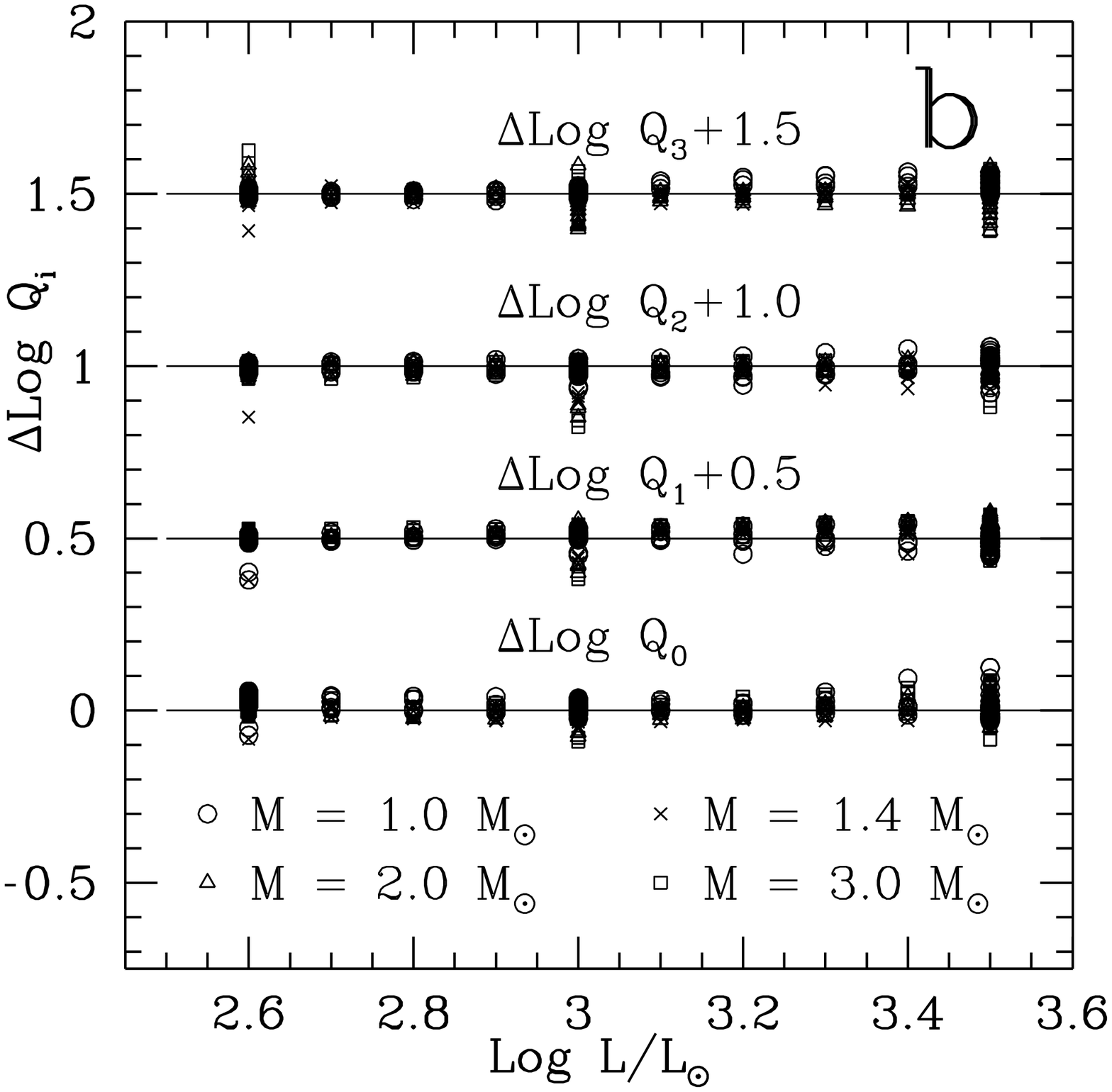}
\caption{The deviations of the analytic fitting (eq.~[\ref{eq21}] in
the text) of the pulsation constants from the values given by
calculatins of oscillations, $\log Q_i=\log Q_i-\log Q_{i, ap}$,
(i=0--3), a) as a function of effective temperature; and b) as a
function of luminosity. solid dots for M=1M$_\odot$, plus signs
1.4M$_\odot$, triangles 2.0M$_\odot$ and squares
3.0M$_\odot$.}\label{fig5}
\end{figure*}

\begin{table}
\caption{coefficients in Eq.(\ref{eq21})}\label{table2}
\begin{tabular}{crrrr}
\hline
F & mode & 1th & 2th & 3th\\
\hline
 a0 & -1.193 & -1.451 & -1.625 & -1.728\\
 a1 & +0.80 & -0.69 & -0.80 & -0.60\\
 a2 & +2.0 & +0.34 & +0.20 & -0.18\\
 a3 & -45  & +17 & +20 & +7.0\\
 b0 & -0.19 & -0.075 & +0.05 & 0.0\\
 b1 & -2.90 & +0.96 & +0.55 & +0.65\\
 b2 & -2.75 & +0.28 & -5.0 & -4.0\\
 b3 & +110  & -28 & 0.0 & 0.0\\
 c0 & +0.34 & -0.015 & -0.04 & +0.01\\
 c1 & +0.70 & -0.10 & -0.17 & -0.49\\
 c2 & -10 & +1.3 & +3.0 & +1.7\\
 c3 & -50 & 0.0 & 0.0 & +10\\
 d0 & -0.80 & +0.15 & +0.06 & -0.08\\
 d1 & 0.00 & +0.30 & +0.75 & -0.89\\
 d2 & +45 & +4.0 & -4.4 & -3.2\\
 d3 & +50 & 0.0 & 0.0 & +90\\
 e0 & +0.18 & -0.06 & +0.04 & +0.06\\
 e1 & -3.5 & 0.0 & +0.27 & 0.0\\
 e2 & -21 & 0.0 & 0.0 & 0.0\\
 e3 & +190 & 0.0 & 0.0 & 0.0\\
 f0 & -0.36 & +0.05 & -0.37 & -0.32\\
 f1 & +12 & -1.0 & -1.7 & -1.3\\
 f2 & +59 & 0.0 & +16 & +22\\
 f3 & -650 & 0.0 & 0.0 & -650\\
\hline
\end{tabular}
\end{table}

The coefficients in above expressions, $(a-f)_n$, from the
fundamental to the 3rd overtone are given in table~\ref{table2}.
Figs.~\ref{fig5}a and \ref{fig5}b demonstrates the difference
$\Delta\log Q_i=\log Q_i-\log Q_{i,ap}$ (i=0-3) between the calculated pulsation
constants $Q_i$ for the nearly 600 models and their analytic fitting
$Q_{i,ap}$ as functions of effective temperature $\log T_e$ (\ref{fig5}a)
and luminosty (\ref{fig5}b). It is clear from fig.~\ref{fig5} that a
pretty good fitting is secured in the domain defined by $1\leq
M/M_\odot\leq 3$, $3.4< \log T_e <3.7$ and $2.6< \log\L/L_\odot
3.5$.

\section{The excitation mechanism for the red giant variables}

By using a work diagram, the excitation and damping on the
oscillations of stars due to different factors can be clearly
illustrated and analysed quantitatively. The definition and derivation of the
integrated work (IW) can be found in the Appendix. The total IW
$W_{all}$ is composed of the following 3 components,

\bq W_{all}=W_P+W_{Pt}+W_{vis} \label{eq22}\eq

where $W_P$ is the pressure component (including radiative
pressure), $W_{Pt}$ is the turbulent pressure one, while $W_{vis}$
is that of turbulent viscosity, all being functions of depth ($M_r$
or $\log P$). In our work, the IW is counted from the stellar center
and integrated outwards, therefore the total IW $W_{all}$ at the
surface should be the growth rate of pulsation amplitude
$\eta=-2\pi\omega_i/\omega_r$, where $\omega_i$ and $\omega_r$ are
respectively the real imaginary and real part of the complex
pulsation frequency $\omega=i\omega_i+\omega_r$, i.e.,

\bq \eta=W_{all}\left(M_0\right)\label{eq23}\eq

When $W_{all}(M_r)$ (or its one component) increases outwards,
this region is therefore a pulsational excitation zone (of
the corresponding component), otherwise it will be a damping zone.

\begin{figure}
\includegraphics[width=84mm]{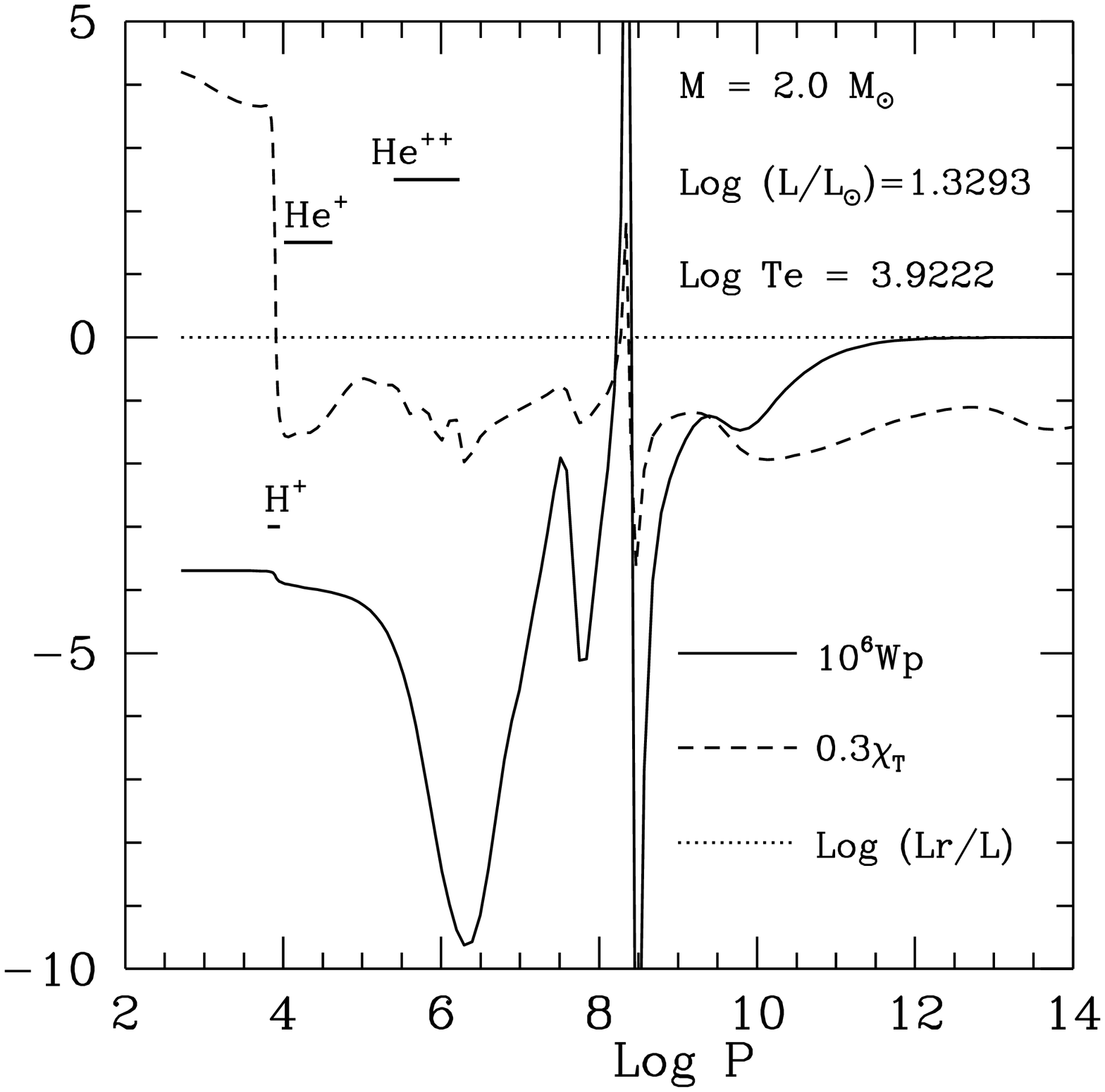}
\caption{The integrated work (IW) of the fundamental mode for a hot
star located outside the blue boundary of the $\delta$ Scuti
instability strip. The solid line is the integrated work (IW),
dashed line the opacity, and dotted line the fractional radiation
flux $L_r/L$ as functions of depth ($\log P$. The 3 heavy horizontal
line on the upper left part of the figure represent the locations of
the ionization zones of hydrogen and the two of helium.}\label{fig6}
\end{figure}

It is well known that the quasi-Cepheid variables (including Cepheids of
populations I and II, RR Lyrae and $\delta$ Scuti stars. this term
will be used hereafter) are excited by the $\kappa$ mechanism.
The existence of the blue edge of the instability strip is due to the fact,
when stellar surface temperature is too high, the ionization zones
of hydrogen and helium that is responsible for $\kappa$ mechanism
are very near the surface, their mass and heat capacity are too small,
and they produce far less excitation that is not enough to
compensate the damping at the deep interior, therefore oscillations
cannot excited for these hot stars. This is shown by the plot of IW
versus depth in fig.~\ref{fig6} for a hot star ($M=2.0M_\odot$,
$\log L/L_\odot$ and $\log T_e=3.9222$) located outside the blue edge
of the $\delta$ Scuti instability strip. It can be seen from
fig.~\ref{fig6} that convection in the whole envelope is negligible.
In the deep interiors below the second ionization zone of
helium, the overall trend of the integrated work $W_P$ is decreasing
towards the surface of ster due to the radiation damping. Near $\log P=8$,
the $W_P$ curve shows  an abrupt ascending followed by a quick
descending, this is due to the iron absorption peak over there. For
$\log P\sim 6.0-4.0$, $W_P$ increases quickly, and it is excited by
the $\kappa$ mechanism in the first and second ionization zones of helium.
There is a little rise up at $\log P\sim 4.0$ which is made by the
$\kappa$ mechanism due to ionization of hydrogen. Nevertheless, all
the excitation in the ionization zones of hydrogen and helium are
not enough to compensate the radiation damping arise from the deep
interiors, this makes $W_P<0$ for this model, therefore the star is
pulsationally stable.

\begin{figure}
\includegraphics[width=84mm]{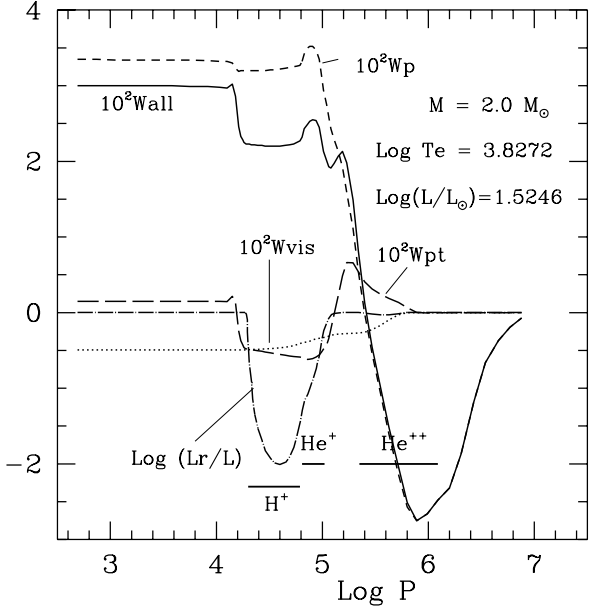}
\caption{The integrated work (IW) of the fundamental mode for a warm
star within the $\delta$ Scuti instability strip as a function of
depth (the solid line). The runs of $W_P$ (the dashed line, gas
pressure component), $P_t$ (the long dashed line, turbulent pressure
component), $W_{vis}$ (dotted line, turbulent viscosity component),
and $\log L_r/L$ (dot-dashed line) are also shown. The parameters
including mass, luminosity and effective temperature of the star are
labeled. The coupling between convection and oscillations has been
taken into account.}\label{fig7}
\end{figure}

The ionization zones of hydrogen and helium go deeper inside with
decrease of effective temperatures of stars, their total mass and
heat capacity become larger, therefore excitation gets enhanced.
When the excitation takes over the radiation damping in the deep
interior, stars will become pulsationally unstable. Fig.~\ref{fig7}
depicts IW versus depth for a warm star within the $\delta$ Scuti
instability strip ($M=2.0M_\odot$, $\log L/L_odot=1.5246$ and $\log
T_e=3.8272$). This star possesses a relative shallow convective
zones of the hydrogen and first helium ionization The second
ionization zone of helium is radiative in this case. Around $\log
P=5.5$ both $W_P$ and $W_{all}$ increase quickly outwards due to the
exciting of the second ionization zone of helium. The convection
component ($W_{Pt}$ and $W_{vis}$) is much smaller than the
excitation due to radiation. This excitation is also greater than
the radiation damping in the deep interiors, therefore the star
become unstable.

\begin{figure*}
\includegraphics[width=84mm]{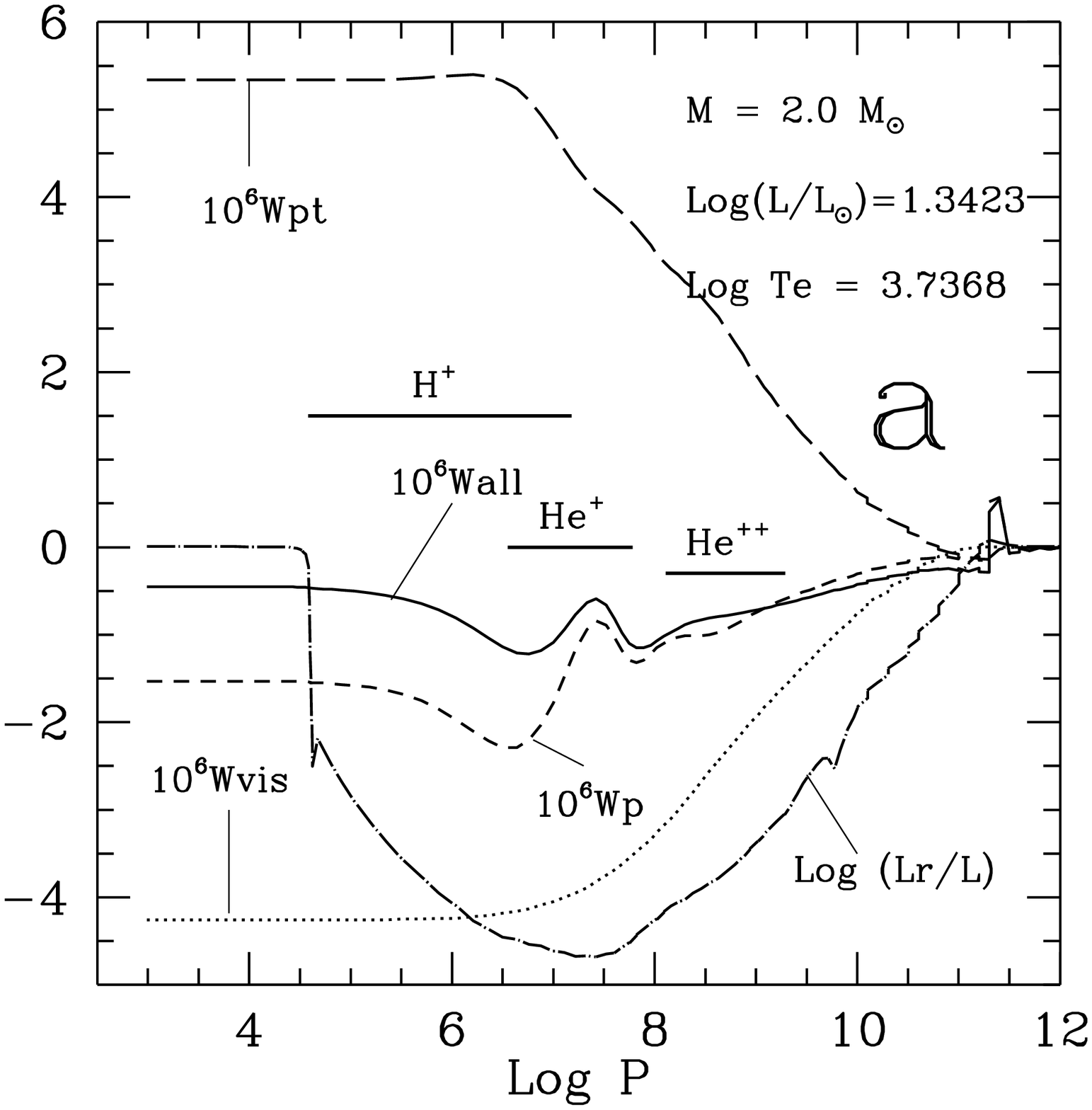}\includegraphics[width=84mm]{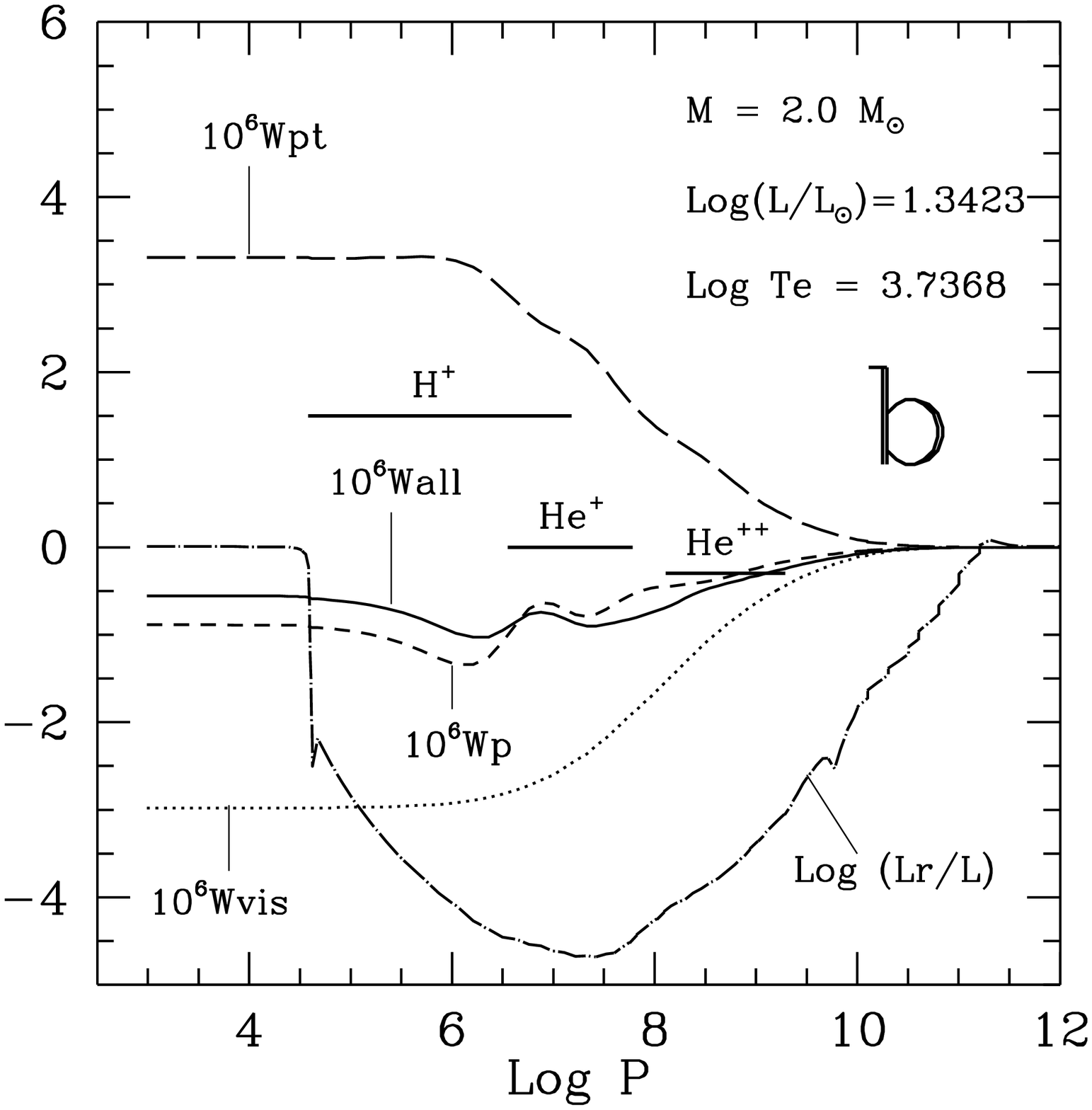}
\caption{Similar to fig.~\ref{fig7}, but for the integrated work
(IW) as a function of depth for a pulsationally stable star located
in between the $\delta$ Scuti instability strip and the PRG
instability strip. a) the fundamental mode; b) the 1st
overtone.}\label{fig8}
\end{figure*}

Going to even lower effective temperatures, convection zone in stars
becomes deeper. When the second ionization zone of helium becomes
complete convective, the excitation due to radiation
$\kappa$-mechanism reduced largely. In the deep interior, however,
the thermodynamic coupling between convection and oscillations works
now as a damping for pulsation that makes the star stable, this is
true at least for the fundamental and the first overtone. The IW for
a yellow giant ($M=2.0M_\odot$, $\log L/L_\odot=1.3426$ and $\log
T_e=3.7368$), which locates in between the $\delta$ Scuti
and the pulsating red giant instability trips, is shown in
fig.~\ref{fig8}. Our calculations demonstrate that this star is
stable at least in the first 7 low order overtones.

\begin{figure*}
\includegraphics[width=84mm]{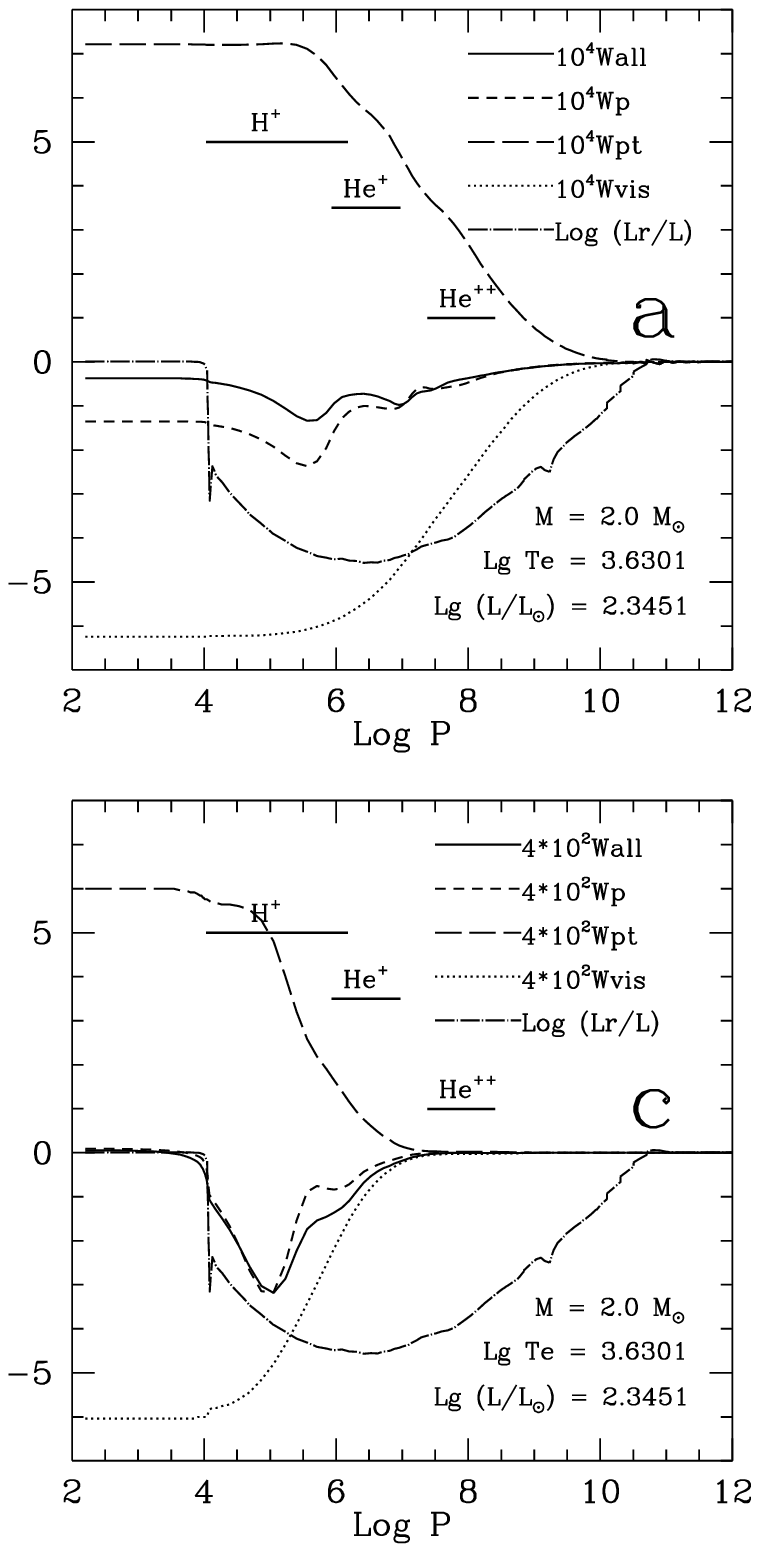}\includegraphics[width=84mm]{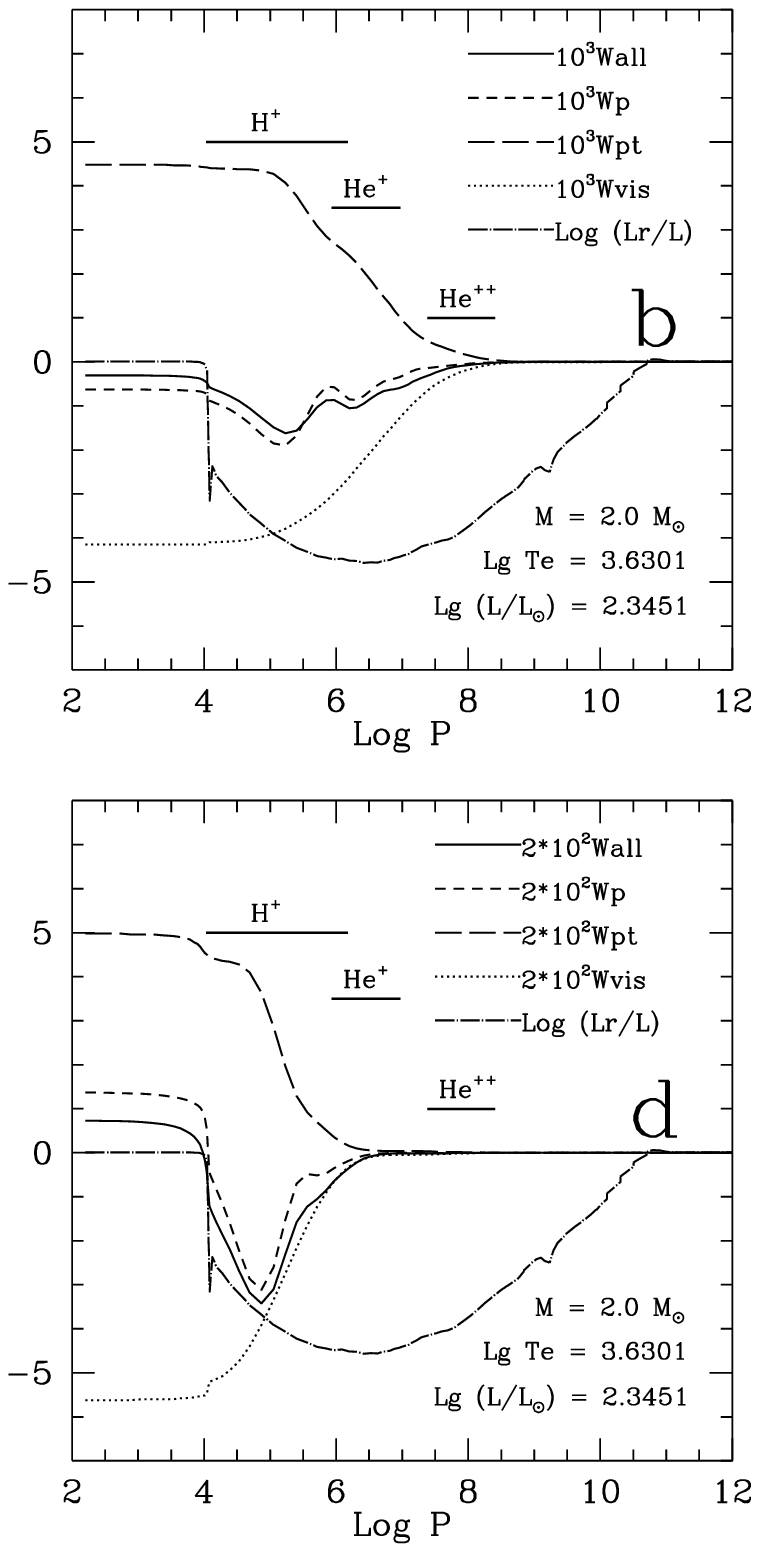}
\caption{Similar to fig.~\ref{fig7}, but for a RGB star. a) the
fundamental mode; b) the 1st overtone; c). the 2nd overtone; d) the
3rd overtone.}\label{fig9}
\end{figure*}

With even lower effective temperature and higher luminosity,
pulsation instability moves to lower order modes. In
figs.~\ref{fig9}a--\ref{fig9}d shows IW of the fundamental and the
first 3 overtone modes versus depth for a star on the RGB
($M=2.0M_\odot$, $\log L/L_\odot=2.3451$ and $\log T_e=3.6301$). For
red giant stars, the pulsation amplitude decreases very quickly
going downwards from the surface, and tends to vanish at the center.
The effective pulsation region can be seen directly from
the IW plot, in which the radiative flux is much smaller than the
convective one. This infers that the gas pressure component of IW,
$W_P$, mainly comes from the thermodynamic coupling between
convection and oscillations. In the deep interior ($\log P\geq 5$),
the general trend of $W_P$ is to decrease with decreasing $\log P$,
meaning that the thermodynamic coupling between convection and
oscillations functions as a damping. Near to the stellar
surface, $W_P$ increases outwards, that is resulted from the radiative
modulation excitation in the radiative flux gradient zone (Xiong,
Cheng \& Deng 1998). It is clear from figs.~\ref{fig9}a--\ref{fig9}b
that, for these red giant stars, the effects of radiation and the
thermodynamic coupling between convection and oscillations ($W_P$) are
already much smaller than that of dynamical coupling ($W_{Pt}$ and
$W_{vis}$). For red giants with intermediate to low luminosity,
$W_{Pt}$ ($>$0) and $W_{vis}$ ($<0$) have similar absolute value while
having opposite signs. The fundamental and the first 2 overtone
modes have $W_{all}<0$, therefore are pulsationally stable; for
modes higher than the 3rd overtone, however, $W_{all}>0$ therefore
the stars become unstable. As shown in table~\ref{table1}, all the
low order modes are stable for low luminosity red giants which are
unstable only at the high order overtones. Going to stars of higher
luminosity and lower effective temperature, pulsation instability
shifts steadily towards lower order modes, the pulsational amplitude
growth rates increase very quickly. For the luminous Mira variables
the fundamental mode and the low oder overtones are only linear
unstable modes, all the modes above 4th order become stable (Xiong,
Deng \& Cheng 1998). Such a trend is marginally shown in
fig.~\ref{fig1} because only the first 5 pulsation modes are
displayed, and no pulsation stability coefficientscan be given.
Table~\ref{table1} is can make up this shortcoming, which shows that all
modes below 8th overtone are pulsationally stable for all population
I stars with $M=1.0M_\odot$ and $\log T\geq 3.70$.  It is clearly shown
from table~\ref{table1} that the pulsation instability shifts
towards lower order modes and amplitude growth rates
increase with increase of luminosity and decrease of effective
temperature of stars. Due to numerical difficulties, we unfortunately
could not construct equilibrium model for high luminosity red giants
under complete non-local treatment of convection. The most luminous
and coolest model got by us is No.40 in the table. However, the general
trend mentioned above can be seen confidently from the table~\ref{table1}
and fig.~\ref{fig1}. This trend agrees well with our previous
work on the luminous long period variables (Xiong, Deng \& Cheng 1998).
The only difference
between previous and the current works is the constructing of the
equilibrium model, the non-adiabatic oscillations are calculated in
exactly the same way: both are treated with our non-local convection
theory. As discussed in the previous section, the amplitude growth
rates of SARVs are smaller than those of luminous Miras
by 1--3 orders of magnitude, very special care is required in
the current work where all the static equilibrium models for the
non-adiabatic oscillations of SARVs are computed using our
complete non-local convection theory. However, the approximate quasi
non-local convection models are applied in the previous work for
the high luminosity LP variables, which is a modification of the
local convection models. In the convectively unstable zone, all the convective
quantities in the quasi non-local models take the same value as in
the local convection models, while in the overshooting zone, they
are extended from the local models following a analytic asymptote of
the properties of overshooting outside the unstable zone. Comparing
the numerical calculations between local and non-local models,
it shows that within a convectively unstable zone, they are
rather close to each other, with the sensible difference only near
the boundaries and in the adjacent overshooting zones (Xiong 1989b).
For luminous red variables having very large pulsation amplitude
growth rates, such a approximate treatment may induce a certain
amount of uncertainty to value of the growth rates, however, the
qualitative results on the stability of these stars will not be
affected. The numarical tests confirm this inference.

By the way, the so-called frozen convection approximation is still
favored by some workers (Gong et al. 1995) in the study of non-adiabatic
oscillations of stars. They assume that convection is unchanged
during the course of stellar oscillations. Such an approximation is
correct only when the pulsation timescale is much smaller than that
of convective motion. Unfortunately, these timescales are usually
the same order of magnitude within the convective zones. Therefore
it is clear that this kind of treatment must be
incorrect in most cases, especially for analyzing the instability of
the low temperature stars with extended convective envelopes. The
only chance for the frozen convection approximation to be reasonable
is for the study of the blue edge of the instability strip, because
for hot star near the blue edge, convection zone is rather shallow so
that radiation dominates, and the coupling between convection and
oscillations has little influence on the stability of these objects.
In any case, this approximation can explain neither the red edge of
the Cepheid instability strip, nor the blue edge of the pulsating
red giant instability strip. We would like to conclude
confidently that if the coupling between convection and oscillations
is excluded, namely the frozen convection approximation is applied,
all stars located to the red side of the blue edge of the Cepheid
instability strip in the HRD will be pulsationally unstable.

\begin{figure*}
\includegraphics[width=84mm]{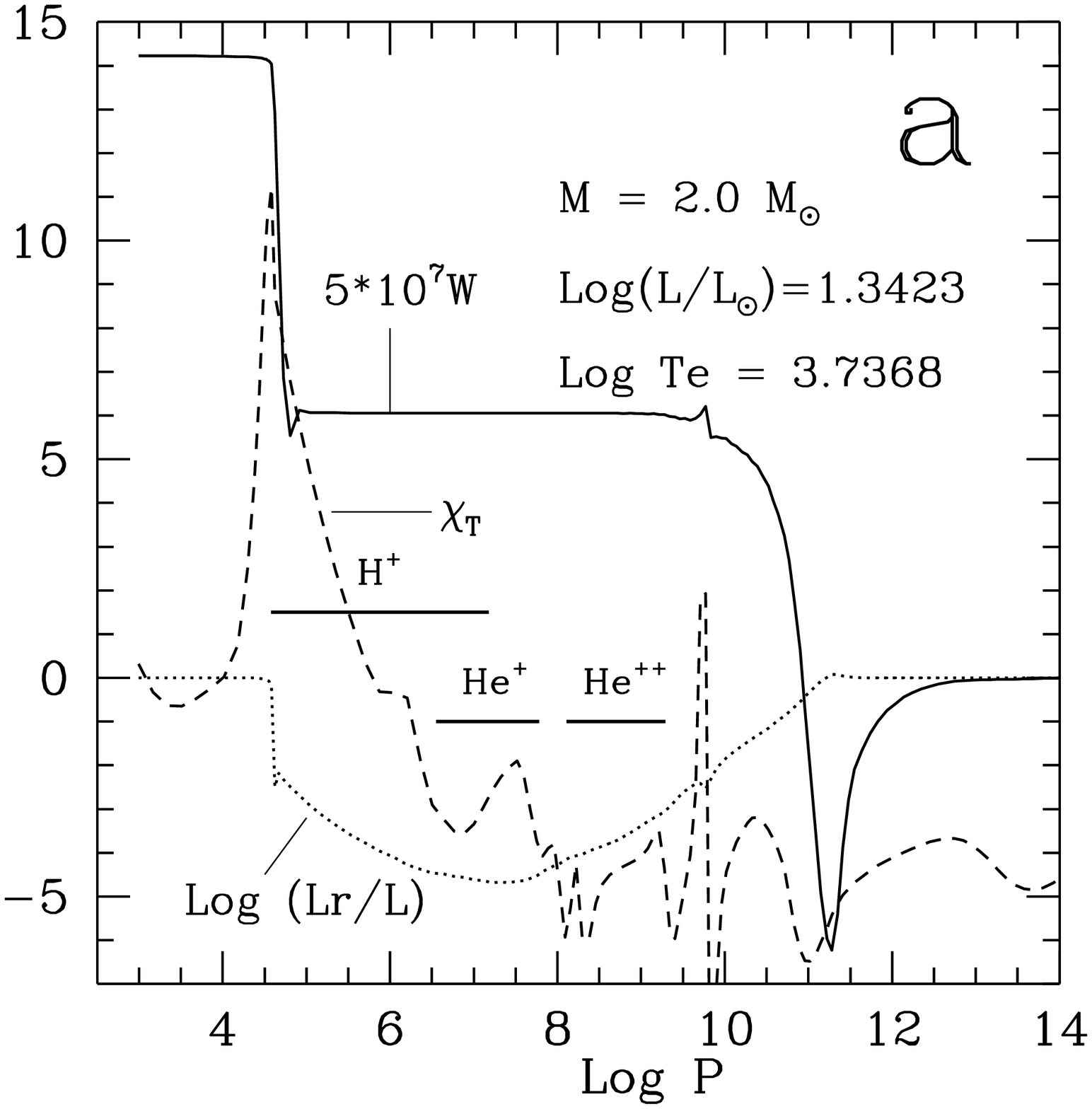}\includegraphics[width=84mm]{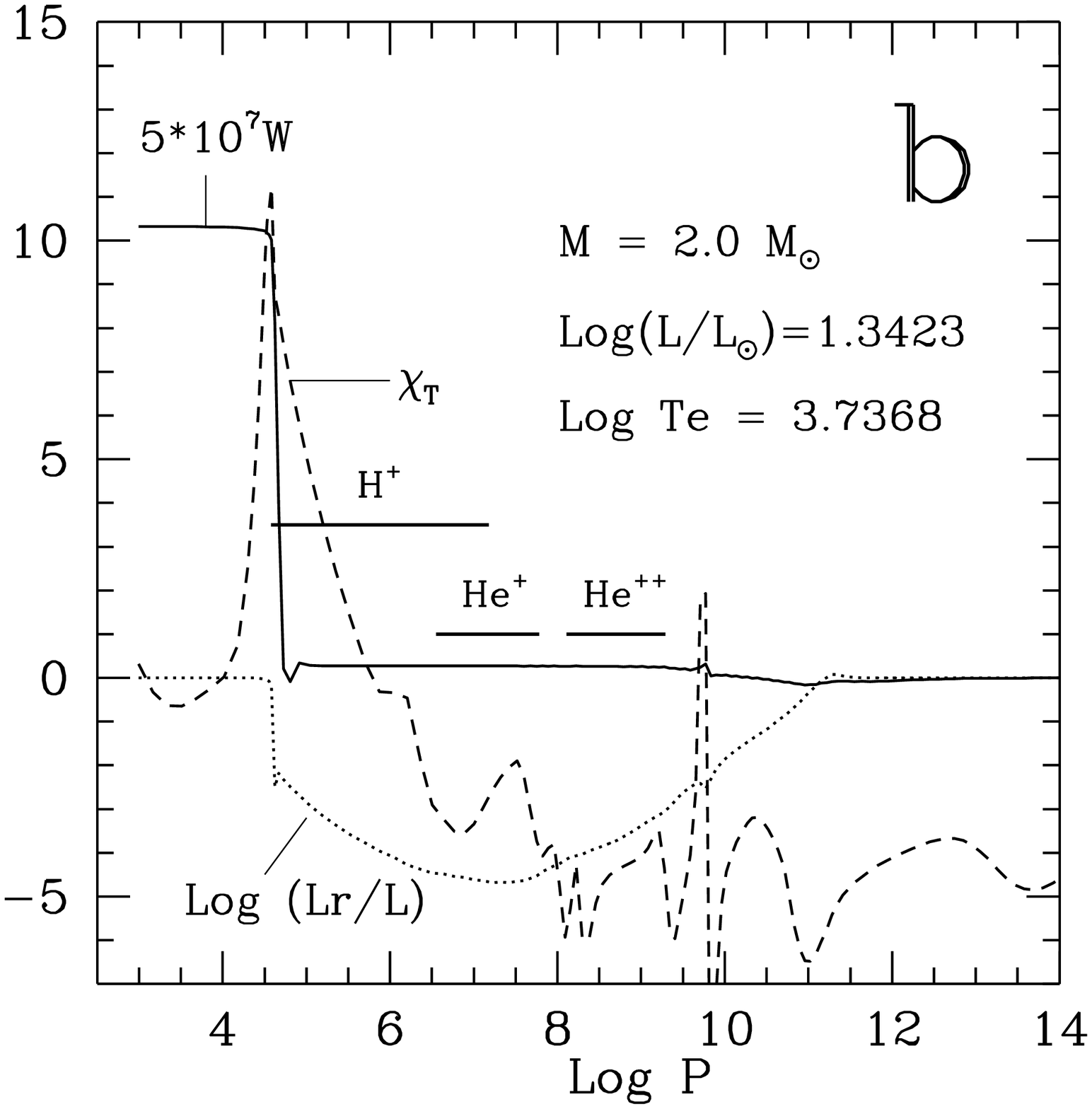}
\caption{The integrated work (IW) as a function of depth for the
same star as fig.~\ref{fig8}, but without taking into account the
coupling between convection and oscillations. a) the fundamental
mode; b) the 1st overtone.}\label{fig10}
\end{figure*}

Figs.~\ref{fig10}a and \ref{fig10}b display the IW versus depth for
the fundamental and the 1st overtone of the same stable star as
in fig.~\ref{fig8}, which is located in between the $\delta$ Scuti
and the red giant instability strips, but the coupling between
convection and oscillations is not considered in this calculation. In
fig.~\ref{fig10}a, there are two rapid increase of IW in the surface
and bottom boundaries of the convection zone. They are resulted from
the radiative modulation excitations produced in the radiative flux
gradient regions. Although being a result of radiative flux and
related to opacity, the radiative modulation excitation is
completely distinct from the radiative $\kappa$ mechanism. The
former one exists only in the radiative flux gradient region whose
functionality can be found in our previous work (Xiong, Cheng \&
Deng 1998). Obviously, the ionization zones of hydrogen and helium
where $\kappa$ mechanism operates, are now fully convective, in that the
radiative flux is only a tiny fraction of the total flux.  As shown
in fig.~\ref{fig10}a, $W_P$ is virtually a horizontal line in the
ionization zones of hydrogen and helium. In the deeper interior,
radiation ought to be a damping, but in the radiative flux gradient
region just inside the
lower boundary of the convective zone, $W_P$ shows a abrupt jump. This
is a good evidence to show that the radiative modulation excitation
is not the usual $\kappa$ mechanism. For the 1st overtone
(fig.~\ref{fig10}b), $W_P$ shows only a single abrupt jump inside the
upper boundary of the convective zone. This is due to the fact that
oscillations tend to be in further outer layer for the 1st overtone
than the fundamental mode, near the lower boundary of convective
zone, the pulsation amplitude is already very small, the radiative
modulation excitation is therefore much less.

Figs \ref{fig8}a and \ref{fig8}b show the IW versus depth for the
same stellar model as the one described in above paragraph, but
having the coupling between convection and oscillations taken into
account. There is no any similarity in the shapes of IW shown in
fig.~\ref{fig8} and fig.~\ref{fig10}. In fig.~\ref{fig8}, the
absolute values of $W_{Pt}$ and $W_{vis}$ are both much greater than
that of $W_P$, and the $W_P$ curve runs completely in a different
form from that in fig.~\ref{fig10}. This difference is caused by
coupling between convection and oscillations. The effects of
radiation have aready overwhelmed by that of convection
for these low temperature stars.

\section{Conclusions and discussion}

In this work, we have studied the linear non-adiabatic oscillations
of the evolutionary models with $M=1-3M_\odot$ using a non-local
time-dependent theory of convection. The excitation mechanism for
the instabilities are studied. The main results can be concluded in
the following,

\begin{enumerate}

\item Two separate pulsational instability strips are found in the
HRD: the hotter on is the $\delta$ Scuti instability strip, while the
cooler one is the pulsating red giant instability strip. The two strips are
intercepted by a region of pulsationally stable (at least from the
fundamental up to the 4th overtone) yellow giants;

\item The $\delta$ Scuti stars and the pulsating red giants (PRGs)
are two distinct types of variables. The former one is excited by
the radiative $\kappa$ mechanism and the later one is drived by the
coupling between convection and oscillations;

\item For the low luminosity PRGs, pulsations are in the high order
modes. Going to lower effective temperature and higher luminosity,
pulsationally unstable modes shift gradually to lower orders
from higher orders;

\item In the deep interior of convective zone far away from the
boundary, generally speaking, the thermodynamic coupling between
convection and oscillations is a damping against pulsation
This is the true reason for the existence of the red edge of the
Cepheid instability strip. Roughly speaking, when
convection zone extend below the second ionization zone of helium,
the red edge of Cepheid instability strip shows up;

\item Obviously, turbulent viscosity is always a damping against
oscillations, whereas turbulent pressure is a excitation mechanism.
The dynamical coupling between convection and oscillations plays a
key role in exciting the oscillations of PRGs;

\item For the instability of the low temperature stars having
extended convective zone, the coupling between convection and
oscillations is absolutely a dominant factor. When the coupling
is excluded, neither the red  edge of the Cepheid instability
strip nor the blue edges of the PRGs instability strip can be
explained, and all stars on the right side of the red side of
the blue edge of the Cepheid instability strip will be
pulsationally unstable;

\item The theoretical Peterson diagram given in this work is very
similar to that of observations by Soszy\'nski et al. (2004).
However, based on our calculations of period ratios, we would like
to guess that very likely the P4 in their work is in fact the
5th overtone, while P3 is a combination of the 3rd and 4th
overtones. For the observed modes with period ratio close to 1, no
explanation can be made at this point.

\end{enumerate}

PRVs are a group of low temperature red
giant stars located outside the Cepheid instability strip in the
HRD. Their exciting mechanism was not known in the past. Based on
a non-local time-dependent theory of convection, the present work
provides a natural solution that explains 2 types of variables of
distinct properties: $\delta$ Scuti stars (and all the Cepheid like
variables) and the PRGs. In previous section, we have addressed the
existence of the red edge of the Cepheid instability strip, and the
presence of another PRV instability strip at a lower temperature
region in the HRD outside the Cepheids instability strip.

In the {\it General Catalogue of Variable Stars}(Kholopov et al.
1985--1990), the long period variables (LPVs) are classified as 3
primary types: Miras, semi-regulars (SRs) and iiregulars (L).
Following the statistical study of PRVs in the Galactic field and
stellar group (stellar clusters and associations), Eggen (1973a,b,
1975, 1977) made another classification according to pulsation
amplitude, which is again in 3 major types: Large-amplitude red
variables (LARVs: $\Delta V>2$mag.), medium-amplitude red variables
(MARVs: $0.3<\Delta V<2$mag.) and small amplitude red variables
(SARV: $\Delta V\leq 0.5$mag.), where the ones with visual amplitude
less than 0.2 mag. and very short period (20--40days) may form a
individual group. Based on the MACHO database, Wood (2000)
discovered that the PRGs in LMC displayed 5 ridges (or sequences) of
A, B, C, D and E in the $K-\log P$ diagram. The Miras, located in
sequence C, have the largest amplitude and for sure represent the
fundamental mode. Stars falling into sequence B are SRs possessing
smaller amplitudes. In sequence C, one finds the stars with
relatively smaller amplitudes ($\Delta V<1.5$mag), also being SRs.
Stars with very small amplitudes ($\Delta V<0.2$) falling in
sequence A are very likely the SARVs as classified by Eggen.
Sequences B and A may correspond to the 1st--3rd overtones (Wood,
2000). Soszy\'nski et al. (2004, 2005) confirmed the discovery of
Wood using OGLE-II and OGLE-III data, and they further pointed out
that the OSARGs (OGLE small amplitude red giants=SARVs) and the LPVs
(Miras) are two different types of pulsating stars. On the
Period-Luminosty (PL) diagram (periods--W$_I$), the primary periods
of OSARGs fall in sequences A and B, and none of the second and the
third dominant period--W$_I$ relation of OSARGs fall in sequences C;
On the contrary, the primary period-W$_I$ relation fall in sequences
C and B, and none of the second and the third dominant period--W$_I$
relation goes in sequence A. Based on classification as such, in the
PL diagram of Wood's ridges A and B of OSARGs, can be divided into 4
separate ridges a$_1$, a$_2$, a$_3$ and a$_4$ for OSARGs brighter
than the tip of RGB, and b$_1$, b$_2$, b$_3$ and b$_4$ for OSARGs
fainter than the tip of RGB. According to the theoretical Peterson
diagram, it is confident to believe that ridge C is the fundamental
model, a$_1$ (b$_1$) and a$_2$ (b$_2$) are respectively the 1st and
the 2nd overtones. Whereas, a$_3$ and a$_4$ are not likely to be
single mode sequence: a$_3$ (b$_3$) is a mixture of the 3rd and the
4th overtones, and a$_4$ (b$_4$) is a mixture of the 4th and the 5th
overtones. Otherwise, as discussed in section~3, it is difficult to
understand the observational Peterson diagram (Soszy\'nski et al.
2004). As shown in fig.~\ref{fig3}, the period ratios are hard
affected by stellar mass, following similar arguments, they are
neither affected by metallicity. Because metallicity determines only
the location of the instability strip and the ML relation. As both
mass and metallicity hardly influence the period ratios, our results
for mode identifications should be reliable.

Observational evidence provided by Wood and Soszy\'nski et al.
demonstrate that there is a tight link between the properties of
PRGs and their oscillation modes. It is clear that primary period of
Miras are in the fundamental mode (Soszy\'nski et al. 2004), the
primary period of SRs are either in the first overtone or the
fundamental mode, the second the third dominant periods of them
should not be higher than the 4th overtone. On the
contrary, most of the of the SARGs are pulsating in the 3rd--4th
overtones ($\sim 50$\%) or in the 2nd overtone ($\sim 20$\%), and
their fundamental mode is pulsationally stable. How to inspect
theoretically such observations?

Based on the theory of oscillations, when neglecting both rotation
and magnetic field, the pulsation properties of stars depend
primarily on luminosity and effective temperature, and secondarily
on mass and metallicity. It follows from table~\ref{table1} that,
for the faint RGs, pulsation instability starts in high order
overtones, and it shifts to lower orders with increase of luminosity
and decrease of effective temperature, while the pulsation amplitudes also
grows. By comparing the pulsational instability for different
stellar masses, as shown in fig.~\ref{fig1}, going towards higher
masses, the blue edge of the instability shifts to higher luminosity
and higher temperature. This theoretical results agree well the
observations. Eggen (1977) studied the distribution in
M$_{bol}$--(R-I) diagram for PRVs in the Galactic field and stellar groups,
and found out that the SARVs are indeed located in the high
temperature ($0.8\leq (R-I)_0\leq 1.6$) and low luminosity region,
the LARVs stay in the low temperature and high luminosity area,
whereas MARVs stay in the intermediate region. Henry et al. (2000)
also shown that all the K5--M0 giants are SARGs, they are located in
the high temperature and low luminosity region in the HRD. Percy \&
Parkes (1998) studied the pulsation modes for 13 SARGs with
well-determined period and accurate Hipparcos parallaxes, and
claimed that 9 of the stars are pulsating in the 1st or the 2nd
overtone, 1 in the 3rd overtone, while only 3 in the fundamental or
the 1st overtone. Soszy\'nski et al.'s observational study shown
that non of all their OSARGs is oscillating in the fundamental mode.
Therefore, it can be conclude that our theoretical predictions agree
pretty well the observational trend. However, it is difficult to
understand that, on the $\log P$--M$_I$ diagram of Soszy\'nski et
al. (2004), the OSARGs extends almost all the way to the brightest
end. We have not yet understant OSARGs. Why they are so different
from the Mira and SR variables at the same luminosity? It can be
seen from fig.~\ref{fig1} that the blue edge of PRGs instability
stip move toward higher temperature and luminosity with increase
of stellar mass. OSARGs, we suppose, are more massive and hotter
than Mira and SR variables with the same luminosity. The brightest
OSARGs may be a group of younger massive red giants. If this guess
is true, there shoud not be lumilous OSARGs in the old clusters.

\section*{Acknowledgements}
The Chinese National Natural Science Foundation (CNNSF) is
acknowledged for support through grants 10573022, 10173013, 10273021
and 10333060.

\appendix
\section{The integrated work}

Using the eq.~\ref{eq2} in the text, its linear pulsation equations
can be give as,

\bqa\lefteqn{4\pi r^3\frac{d}{dM_r}\left(\delta P+\delta
P_t\right)}\nonumber \\
  & & \mbox +{d\over{dM_r}}\left\{4\pi r^3\rho \Pi^{11}\left[
  \frac{\delta \Pi^{11}}{\Pi^{11}}-\frac{d}{d\ln r}\left({{\delta r}\over r}
  \right)\right]\right\}\nonumber\\
  & & \mbox -\left[r^2\omega^2+4\frac{GM_r}{r}
  +\frac{d}{dM_r}\left(4\pi r^3\rho \Pi^{11}\right)\right]\frac{\delta r}{r}=0,
  \label{eqA1}
\eqa

where $P_t=\rho\chi^2$ is the turbulent pressure, $\omega$ is the
complex frequency of oscillations defined as,

\bq \omega=\omega_r+i\omega_i,\label{eqA2}\eq

putting (\ref{eqA2}) into (\ref{eqA1}), solving the equation for
$\omega$, we have,

\bqa\lefteqn{r^2\left(\omega^2_r+2i\omega_i\omega_r-\omega^2_i\right)
\frac{\delta r}{r}}\nonumber \\
  & &\mbox =4\pi r^3\frac{d}{dM_r}\left(\delta P+\delta
  P_t\right)\nonumber \\
  & & \mbox +\frac{d}{dM_r}\left\{4\pi r^3\rho \Pi^{11}
  \left[\frac{\delta \Pi^{11}}{\Pi^{11}}-\frac{d}{d\ln r}
  \left({{\delta r}\over r}\right)\right]\right\}\nonumber \\
  & &\mbox -\left[4\frac{GM_r}{r}+3\frac{d}{dM_r}
  \left(4\pi r^3\rho \Pi^{11}\right)\right]{{\delta r}\over r},\label{eqA3}
\eqa

The next step is multiplying eq.~(\ref{eqA3}) with $\pi \left(\delta
r^*/r\right)dM_r$, then integrating it for $dM_r$ from the center to
the surface. Baring in mind that
the relative pulsational amplitude at the center is zero and the
pressure and density vanish at the surface, and Separating the real and
imaginary parts, we have,

\bqa\lefteqn{2\pi\omega^2_r\frac{\omega_i}{\omega_r}\int^{M_0}_0\delta
r\delta r^*dM_r}\nonumber \\
  & &\mbox =4\pi^2 \int^{M_0}_0I_m\left\{r^3\frac{d}{dM_r}
  \left(\delta P+\delta P_t\right)\right.\nonumber \\
  & &\mbox +\frac{d}{dM_r}\left.\left[\rho r^3\Pi^{11}
  \left(\frac{\delta \Pi^{11}}{\Pi^{11}}-\frac{d}{d\ln r}\left(\frac{\delta
  r}{r}\right)\right)\right]\right\}
  \frac{\delta r^*}{r}dM_r\nonumber \\
  & &\mbox =\pi\int^{M_0}_0I_m \left[\left( \delta P+\delta
  P_t\right)
  \frac{\delta\rho^*}{\rho}\right.\nonumber\\
  & &\mbox -\left.\delta \Pi^{11}\frac{d}{d\ln r}
  \left(\frac{\delta r^*}{r}\right)\right]dM_r,\label{eqA4} \eqa

where $I_m$ means to take the imaginary of the term in the brackets.
$-2\pi\omega_i/\omega_r$ is the amplitude growth rate of
oscillation within a cycle $\eta$,

\bqa\lefteqn{\eta=
  -2\pi\frac{\omega_i}{\omega_r}=
  -{\pi\over{E_k}}\int^{M_0}_0I_m\left\{\left(\delta P+\delta P_t\right)
  \frac{\delta\rho^*}{\rho}\right.}\nonumber\\
  & & -\delta \Pi^{11}\left.\frac{d}{d\ln r}
  \left(\frac{\delta r^*}{r}\right)\right\}dM_r.\label{eqA5}\eqa

where

\bq E_k=\omega^2_r\int\delta r\delta r^*dM_r.\label{eqA6}\eq

As from eq.~(\ref{eqA5}), $\eta$ can be divided into a gas pressure
component and a Renold's stress component. We can define the IWs as
functions of depth, $W_P(M_r)$ and $W_{dyn}(M_r)$, to describe the
contributions of the two components,

\bq W_P\left(M_r\right)=-\frac{\pi}{E_k}\int^{M_r}_0I_m\left\{
\delta P\delta\rho^*/\rho^2\right\}dM_r,\label{eqA7}\eq

\bqa\lefteqn{W_{dyn}\left(M_r\right)=
  -\frac{\pi}{E_k}\int^{M_r}_0I_m\left\{{1\over{\rho^2}}\delta
  P_t\delta\rho^*\right.}\nonumber\\
  & &\mbox -\delta \Pi^{11}\left.\frac{d}{d\ln r}\left(\frac{\delta
  r^*}{r}\right)\right\}dM_r,\label{eqA8}\eqa

The gas component of the IW, $W_P(M_r)$, contains both the
contribution of radiative energy transfer and the convective one.
Starting from the energy equation in the text, eq.~\ref{eq3}, we can
have,

\bqa\lefteqn{\delta
  P=\left(\Gamma_1-1\right)\rho\left\{\chi^2\left(\frac{\delta\rho}{\rho}
  -3\frac{\delta\chi}{\chi}\right)\right.}\nonumber\\
  & &\mbox +\left.\frac{1}{i\omega}\left[\delta\epsilon_1
  -\frac{d}{dM_r}\left(\delta L_r+\delta L_c+\delta
  L_t\right)\right]\right\},
  \label{eqA9}\eqa

where $L_r$, $L_c$ and $L_t$ are the luminosities corresponding
respectively to the radiative, the convective and turbulent energy
fluxes. $\epsilon_1$ is the nuclear energy generation rate (with
that of neutrino subtracted). Due to the fact that oscillations drops
off abruptly towards stellar center for giant stars, the calculation
of oscillating stellar model can be simplified by doing the job only
for the models of envelope. Under such a condition, the contribution
of energy generation rate can be safely neglected. Putting
eq.~(\ref{eqA9}) into eq.~(\ref{eqA7}), the contribution of
radiative energy transfer can be singled out from that of convective
energy transfer,

\bq
W_P\left(M_r\right)=W_{Pr}\left(M_r\right)+W_{therm}\left(M_r\right),\label{eqA10}\eq

where,

\bqa\lefteqn{W_{Pr}\left(M_r\right)=}\nonumber\\
  & &\mbox -\frac{\pi}{E_k}\int^{M_0}_0Re\left\{
  \frac{\Gamma_3-1}{\omega}\frac{\delta\rho^*}{\rho}
  \frac{d\left(\delta
  L_r\right)}{dM_r}\right\}dM_r,\label{eqA11}\eqa

\bqa\lefteqn{W_{therm}\left(M_r\right)=\frac{\pi}{E_k}\int^{M_0}_0\left\{
  I_m\left[3\left(\Gamma_3-1\right)\chi^2\frac{\delta\rho^*}{\rho}
  \frac{\delta\chi}{\chi}\right]\right.}\nonumber\\
  & &\mbox -Re\left[\frac{\left(\Gamma_3-1\right)}{\omega}
  \frac{\delta\rho^*}{\rho}\right.\left.\left.\frac{d\left(\delta L_c+\delta L_t\right)}
  {dM_r}\right]\right\}dM_r,\label{eqA12}\eqa

$Re$ means taking the real part of the terms in the brackets.

$W_{Pr}$ is the contribution of radiative energy transfer, while
$W_{therm}$ is that of convective energy transfer, i.e. the
thermodynamic coupling between convection and oscillations.
$W_{dyn}$ is that due to Renold's stress, the dynamic coupling
between convection and oscillations. The second and the third terms
in the dynamic equations (\ref{eq5}) and (\ref{eq8}) for
turbulent Renold stress in the text, represent the turbulent viscosity.
Solving for $\delta P_t$ and $\delta \Pi^{11}$ from the linear
pulsation equations of eqs.~(\ref{eq5}) and (\ref{eq8}), it is not
difficult to isolate the viscosity term which is then put in
eq.~(\ref{eqA8}), and we can have the turbulence viscosity component
of the IW,

\bqa\lefteqn{W_{vis}\left(M_r\right)}\nonumber \\
  & &\mbox =-{\pi\over{E_k}}\int^{M_0}_0I_m\left\{\frac{i\omega\tau_c}
  {{4\over
  3}+i\omega\tau_c}\left[\chi^2\frac{\delta\rho}{\rho}\right.\right.
  \nonumber\\
  & &\mbox +\left.\Pi^{11}\left(\frac{\delta\rho}{\rho}+3\frac{\delta r}{r}\right)\right]
  \frac{\delta\rho^*}{\rho}
  +{4\over 3}\frac{i\omega\tau_{c1}}{1+i\omega\tau_{c1}}
  \left[\chi^2\frac{d}{d\ln r}\left(\frac{\delta r}{r}\right)\right.
  \nonumber\\
  & &\mbox +\left. \Pi^{11}\left(\frac{d}{d\ln r}\left(\frac{\delta r}{r}
  \right)+{3\over 2}\frac{\delta r}{r}\right)\right]\left.\frac{d}{d\ln
  r}\frac{\delta r^*}{r}\right\}dM_r\nonumber\\
  & &\mbox =-\frac{\pi}{E_r}\int^{M_0}_0\left\{{4\over 3}
  \left(\chi^2+\Pi^{11}\right)\left[\frac{\omega\tau_c}{\frac{16}{9}+\omega^2\tau_c^2}
  \frac{\delta\rho}{\rho}\frac{\rho^*}{\rho}\right.\right.\nonumber \\
  & &\mbox +\left.\frac{\omega\tau_{c1}}{1+\omega^2\tau_{c1}^2}
  \frac{d}{d\ln r}\left(\frac{\delta r}{r}\right)
  \frac{d}{d\ln r}\left(\frac{\delta
  r^*}{r}\right)\right]\nonumber\\
  & &\mbox +\frac{\omega\tau_{c1}}{1+i\omega^2\tau_{c1}^2}\Pi^{11}
  \frac{d}{d\ln r}\left(\frac{\delta r}{r}\frac{\delta
  r^*}{r}\right)\nonumber\\
  & &\mbox +\left.Re\left[\frac{3\omega\tau_c\Pi^{11}}{{4\over 3}+i\omega\tau_c}
  \frac{\delta
  r}{r}\frac{\delta\rho^*}{\rho}\right]\right\}dM_r,\label{eqA13}
  \eqa

where $\tau_c$ is the dissipation time scale of turbulence, which is
also the inertia time scale of turbulence, expressed as,

\bq \tau_c=\frac{c_1r^2P}{0.78GM_r\rho\chi},\label{eqA14}\eq

\bq \tau_{c1}=\frac{3}{4\left(1+c_3\right)}\tau_c.\label{eqA15}\eq

Subtracting eq.~(\ref{eqA8}) with eq.~(\ref{eqA13}), one has the
turbulent pressure component of the IW, $W_{Pt}$,

\bq
W_{Pt}\left(M_r\right)=W_{dyn}\left(M_r\right)-W_{vis}\left(M_r\right).
\label{eqA16}\eq

Summing up the four components, $W_{Pr}$, $W_{therm}$, $W_{Pt}$ and
$W_{vis}$, we have the total IW,

\bqa
\lefteqn{W_{all}\left(M_r\right)=W_{Pr}\left(M_r\right)+}\nonumber\\
 & &\mbox W_{therm}\left(M_r\right)
+W_{Pt}\left(M_r\right)+W_{vis}\left(M_r\right),\label{eqA17}\eqa

On the right hand side of eq.~(\ref{eqA17}), only the first term is
the contribution of radiative energy transfer, all the other 3 terms
are due to convection, where $W_{therm}$ is the contribution of the
thermodynamic coupling between convection and oscillations, $W_{Pt}$
is that due to turbulent pressure, and $W_{vis}$ is that of
turbulent viscosity. Under such definitions, the linear amplitude
growth rate of oscillations, $\eta$, will equal the surface value of
$W_{all}\left(M_r\right)$,

\bq
\eta=-2\pi\omega_i/\omega_r=W_{all}\left(M_0\right).\label{eqA18}\eq

It is clear from eq. (\ref{eqA13}) that the turbulent viscosity
component $W_{vis}$ of the IW is always negative, therefore it is a
dumping against oscillations. While the effects of the turbulent
pressure component $W_{Pt}$ and turbulent thermal convection
component $W_{therm}$ depend very much on the phase of the
variations of turbulent pressure $P_t$ and turbulent thermal
convection $L_c=4\pi r^2\rho c_PTV$ relative to that of stellar
oscillations ($\delta\rho/\rho$, or $\delta r/r$). It follows from
the dynamic equations of turbulent convection
(eqs.~\ref{eq1}--\ref{eq5}) that, in general, the relative
variations of convection quantities $\delta\chi/\chi$, $\delta Z/Z$,
$\delta V/V$ and $\delta\Pi^{11}/\Pi^{11}$ always lag behind density
variation $\delta\rho/\rho$ (or -$\delta r/r$) inside a convective
zone. Under local convection approximations, this argument was
proved analytically in our previous work (Xiong 1977). More general
oscillation calculations using non-local convection treatment
further proved this statement (Xiong, Cheng \& Deng 1998, Xiong,
Deng \& Cheng 1998). Therefore one can conclude from
eqs.~(\ref{eqA8} \& \ref{eqA12}) that $W_{Pt}>0$ and $W_{therm}<0$
general hold inside a convective zone, i.e. turbulent pressure is a
excitation mechanism for oscillations, while turbulent thermal
convection (the thermodynamic coupling between convection and
oscillations) is a damping one. Such a theoretical analysis agrees
perfectly with what have been shown in
figures~\ref{fig7}--\ref{fig9} for $W_{Pt}$, $W_{therm}$ and
$W_{vis}$.


\begin{thebibliography}{99}
\bibitem{Alcock00}Alcock, C., et al., 2000, ApJ, 541, 734
\bibitem{Alex94}Alexander, D.R. \& Ferguson, J.W., 1994, ApJ, 437, 879
\bibitem{Afonso}Afonso, C., et al., 1999, A\&A, 344, L63
\bibitem{Bal92}Balmforth, N.J., 1992, MNRAS, 255, 603
\bibitem{Bond01}Bond, I.A. et al, 2001, MNRAS, 327, 868
\bibitem{BBC93}Bressan, A., Fagotto, F., Bertelli, G. \& Chiosi, C., 1993, A\&AS,
100, 647
\bibitem{Coini01}Coini, M.-R.L., Marquette, J.-B. \& Loup, C., 2001,
A\&A, 377, 945
\bibitem{Cox91}Cox, A.N. \& Ostlie, D.R., 1991, Ap\&SS, 210, 311
\bibitem{DX2006}Deng, L. \& Xiong, D.R., 2006, ApJ, 643, 426
\bibitem{Edm96}Edmonds, P.D. \& Gilliland, R.L., 1996, ApJ, 464, L157
\bibitem{Eggen69}Eggen, O.J., 1969, ApJ, 158, 225
\bibitem{Eggen72a}Eggen, O.J., 1972a, ApJ, 174, 45
\bibitem{Eggen72b}Eggen, O.J., 1972b, ApJ, 177, 489
\bibitem{Eggen73a}Eggen, O.J., 1973a, ApJ, 180, 857
\bibitem{Eggen73b}Eggen, O.J., 1973b, ApJ, 184, 793
\bibitem{Eggen75}Eggen, O.J., 1975, ApJ, 195, 661
\bibitem{Eggen77}Eggen, O.J., 1977, ApJ, 213, 767
\bibitem{Fox82}Fox, M.W. \& Wood, P.R., 1982, ApJ, 259, 198
\bibitem{Gong95}Gong, Z.G., Li, Y. \& Huang, R.Q., 1995, AcASn, 15,
56
\bibitem{Gough77}Gough, D.A., 1977, ApJ, 214, 196
\bibitem{Henry00}Henry, G.W., Fekel, F.C. \& Henry, S.M., 2000,
ApJs, 130, 201
\bibitem{Hinze75}Hinze, J.O., 1975, Turbulence (McGraw-Hill, New
York)
\bibitem{Johnson66}Johnson, H.L., 1966, ARA\&A, 4, 193
\bibitem{Jor97}Jorissen, A., Mowlavi, N., Sterken, C. \& Manfroid,
J., 1977, A\&A, 324, 578
\bibitem{Kholopov8590}, Kholopov, P.N., et al., 1985--1990, General
Catalogue of Variable Stars, 4th ed. (Moscow State University)
\bibitem{kiss03}Kiss, L.L. \& Bedding, T.R., 2003, MNRAS, 343, L79
\bibitem{kiss04}Kiss, L.L. \& Bedding, T.R., 2004, MNRAS, 347, L83
\bibitem{Noda02}Noda, S. \& takeuti, M., Abe, F. et al., 2002,
MNRAS, 330, 137
\bibitem{Ost86}Ostlie, D.A. \& Cox, A.N., 1986, ApJ, 311, 864
\bibitem{Per98}Percy, J.R. \& Parkes, M., 1998, PASP, 110, 1431
\bibitem{Per01}Percy, J.R., Nyssa, Z. \& Henry, G.W., IBVS, 2001,
No. 5209
\bibitem{per03}Percy, J.R. \& Bakos, A.G., 2003, in Garrison
Festchrift Conf. eds. R.O. Gray, C.J. Corbally \& A.G.D. Philip
(Davis press), p.49
\bibitem{Pojmanski02}Pojmanski, G., 2002, Acta Astron., 52, 397
\bibitem{Rogers92}Rogers, F. \& Iglesias, C.A., 1992, ApJs, 79,
507
\bibitem{Rotta}Rotta, J.C., 1951, Z.Phys., 129, 547
\bibitem{Schultheis04}Schultheis, M., Glass, I.S. \& Coini, M.-R.,
2004, A\&A, 427, 945
\bibitem{Sos04}Soszy\'nski, I., Udalski, A., Kubiak, M., et
al., 2004, AcA, 54, 129
\bibitem{Sos05}Soszy\'nski, I., Udalski, A., Kubiak, M., et
al., 2005, AcA, 55, 331
\bibitem{Wood99}Wood, P.R., Alcock, C., Allsmaa, R.A., et al.,
IAU Symp. 191, Asymptotic Giant Branch Stars, ed. T. Le Bertre,
A. L\'ebre and C. Waelkens (San Francisco: APS), 151
\bibitem{Wood00}Wood, P.R., 2000a, PASA, 17, 18
\bibitem{Woz04}Wo\'zniak, P.R., Willians, S.J., Vestrand, W.T. \&
Gupta, V., 2004, AJ, 128, 2965
\bibitem{Wray04}Wray, J.J., Eyer, L. \& Paczy\'nski, 2004, MNRAS,
349, 1059
\bibitem{X77}Xiong, D.R., 1977, Acta Astron Sinica, 18, 86
\bibitem{X89a}Xiong, D.R., 1989a, A\&A, 209, 126
\bibitem{X89b}Xiong, D.R., 1989b, A\&A, 213, 176
\bibitem{XCD98}Xiong, D.R., Cheng, Q.L. \& Deng, L., 1998, ApJ,
500, 449
\bibitem{XCD00}Xiong, D.R., Cheng, Q.L. \& Deng, 2000,
MNRAS, 319, 1079
\bibitem{XDC98}Xiong, D.R., Deng, L. \& Cheng, Q.L., 1998,
ApJ, 499, 355
\bibitem{XD2001a}Xiong, D.R. \& Deng, L., 2001a, MNRAS, 324, 243
\bibitem{XD2001b}Xiong, D.R. \& Deng, L., 2001b, MNRAS, 327, 1137
\bibitem{XD06a}Xiong, D.R. \& Deng, L., 2006a, AcASn, 47, 256
\bibitem{XD06b}Xiong, D.R. \& Deng, L., 2006b, ChA\&A, 30, 379
\bibitem{YAO93}Yao, B.A. \& Qin, D., 1993, IBVS, No.3920
\bibitem{YZQ93}Yao, B.A., Zhang, C.S., Qin, D., 1993, IBVS, No.3955

\end{thebibliography}
\end{document}